%% file: paper.tex
\documentclass[12pt,oneside, a4paper]{article}

\usepackage{amsmath, amssymb, color, graphicx, amsthm}
\usepackage{newclude}

\usepackage[backref=false, style=authoryear, isbn=false, maxcitenames=2, maxbibnames=99, firstinits=true, dashed=false, uniquelist=false, backend=bibtex]{biblatex}

\bibliography{extracted.bib}

\AtEveryBibitem{
\clearlist{language}
\clearlist{address}
\clearlist{location}
}

\DeclareFieldFormat[article]{volume}{\mkbibbold{#1}}%
\DeclareFieldFormat[article]{pages}{#1}%

\renewbibmacro*{volume+number+eid}{%
  \printfield{volume}%
  \printfield{number}%
  \setunit{\addcomma\space}%
  \printfield{eid}}
\DeclareFieldFormat[article]{number}{\mkbibparens{#1}}

\DefineBibliographyStrings{english}{%
  bibliography = {References},
}

\renewbibmacro{in:}{
  \ifentrytype{article}{}{\printtext{\bibstring{in}\intitlepunct}}}

\usepackage{rotating}
\usepackage{multirow}
\usepackage{diagbox}
\usepackage{cancel}
\usepackage{color, colortbl}
\usepackage[list=on]{caption}
\usepackage{setspace}
\usepackage[margin = 2.1cm]{geometry}
\usepackage{bm}
\usepackage{fancyvrb}
\usepackage{floatrow}
\usepackage{eurosym}

\DeclareFloatFont{tiny}{\tiny}

\usepackage{hyperref}
\hypersetup{
    colorlinks,
    linktocpage,
    citecolor=blue,
    filecolor=blue,
    linkcolor=blue,
    urlcolor=blue
}

\newcommand{\myparskip}{.4em}
\newcolumntype{x}[1]{%
>{\centering\arraybackslash}m{#1}}%

\input{macros}

\graphicspath{{graphics/}}
\renewcommand\cite{\textcite}
\newcommand\citep{\parencite}

\title{Sparse Bayesian Time-Varying Covariance\\[-.6em]Estimation in Many Dimensions\\[-.3em]}
\author{Gregor Kastner\footnote{Department of Finance, Accounting and Statistics, WU Vienna University of Economics and Business, Welthandelsplatz 1/D4/4, 1020 Vienna, Austria, +43 1 31336-5593, gregor.kastner@wu.ac.at}\\[-.6em]
WU Vienna University of Economics and Business, Austria}

\begin{document}

\maketitle


\begin{abstract}
We address the curse of dimensionality in dynamic covariance estimation by modeling the underlying co-volatility dynamics of a time series vector through latent time-varying stochastic factors. The use of a global-local shrinkage prior for the elements of the factor loadings matrix
pulls loadings on superfluous factors towards zero.
To demonstrate the
merits of the proposed framework,
the model is applied to simulated data as well as to daily log-returns of $300$ S\&P 500 members. Our approach yields precise correlation estimates, strong implied minimum variance portfolio performance and superior forecasting accuracy in terms of log predictive scores when compared to typical benchmarks.
\end{abstract}

\textbf{JEL classification:} C32; C51; C58

\textbf{Keywords:} dynamic correlation, factor stochastic volatility, curse of dimensionality, shrinkage, minimum variance portfolio

\section{Introduction}

The joint analysis of hundreds or even thousands of time series exhibiting a potentially time-varying variance-covariance structure has been on numerous research agendas for well over a decade. In the present paper we aim to strike the indispensable balance between the necessary flexibility and parameter parsimony by using a factor stochastic volatility (SV) model in combination with a global-local shrinkage prior.
Our contribution is threefold. First, the proposed approach offers a hybrid cure to the curse of dimensionality by combining \emph{parsimony} (through imposing a factor structure) with \emph{sparsity} (through employing computationally efficient absolutely continuous shrinkage priors on the factor loadings). Second, the efficient construction of posterior simulators allows for conducting Bayesian inference and prediction in very high dimensions via carefully crafted Markov chain Monte Carlo (MCMC) methods made available to end-users through the \proglang{R} \citep{r:r} package \pkg{factorstochvol} \citep{r:fac}. Third, we show that the proposed method is capable of accurately predicting covariance and precision matrices which we asses via statistical and economic forecast evaluation in several simulation studies and an extensive real-world example.

Concerning factor SV modeling, early key references include \cite{har-etal:mul, pit-she:tim, agu-wes:bay} which were later picked up and extended by e.g.~\cite{phi-gli:fac, chi-etal:ana, han:ass, lop-car:fac, nak-wes:dynJFE, zho-etal:bay, ish-omo:por}. While reducing the dimensionality of the problem at hand, models with many factors are still rather rich in parameters. Thus, we further shrink unimportant elements of the factor loadings matrix to zero in an automatic way within a Bayesian framework. This approach is inspired by high-dimensional regression problems where the number of parameters frequently exceeds the size of the data. In particular, we adopt the approach brought forward by \cite{car-dou:spa, gri-bro:inf} who suggest to use a special continuous prior structure -- the Normal-Gamma prior -- on the regression parameters (in our case the factor loadings matrix). This shrinkage prior is a generalization of the Bayesian Lasso \citep{par-cas:bay} and has recently received attention in the econometrics literature \citep{bit-fru:ach, hub-fel:ada}.

Another major issue for such high-dimensional problems is the computational burden that goes along with statistical inference, in particular when joint modeling is attempted instead of multi-step approaches or rolling-window-like estimates.
Suggested solutions include \cite{eng-kel:dyn} who propose an estimator assuming that pairwise correlations are equal at every point in time, \cite{pal-etal:fit} who consider composite likelihood estimation, \cite{gru-wes:gpu} who use a decoupling-recoupling strategy to parallelize estimation (executed on graphical processors), \cite{lop-etal:par} who treat the Cholesky-decomposed covariance matrix within the framework of Bayesian time-varying parameter models, and \cite{oh-pat:mod} who choose a copula-based approach to link separately estimated univariate models.
We propose to use a Gibbs-type sampler which allows to jointly take into account both parameter as well as sampling uncertainty in a finite-sample setup through fully Bayesian inference, thereby enabling inherent uncertainty quantification.
Additionally, this approach allows for fully probabilistic in- and out-of-sample density predictions.

For related work on sparse Bayesian prior distributions in high dimensions, see e.g.~\cite{kau-sch:bay} who use a point mass prior specification for factor loadings in dynamic factor models or \cite{ahe-etal:bay} who use a graphical representation of vector autoregressive models to select sparse graphs. From a mathematical point of view, \cite{pat-etal:pos} investigate posterior contraction rates for a related class of continuous shrinkage priors for static factor models and show excellent performance in terms of posterior rates of convergence with respect to the minimax rate. All of these works, however, assume homoskedasticity and are thus potentially misspecified when applied to financial or economic data. For related methods that take into account heteroskedasticity, see e.g.~\cite{nak-wes:dynJFE, nak-wes:dynBJPS} who
employ a latent thresholding process to enforce time-varying sparsity. Moreover, \cite{zha-etal:dyn} approach this issue via dependence networks, \cite{lod-etal:sel} use stochastic search for model selection, and \cite{bas-etal:tim} use time-varying combinations of dynamic models and equity momentum strategies. These methods are typically very flexible in terms of the dynamics they can capture but are applied to moderate dimensional data only.

We illustrate the merits of our approach through extensive simulation studies and an in-depth financial application using 300 S\&P 500 members.
In simulations, we find considerable evidence that the Normal-Gamma shrinkage prior leads to substantially sparser factor loadings matrices which in turn translate into more precise correlation estimates when compared to the usual Gaussian prior on the loadings.\footnote{Note that in contrast to e.g.\ \cite{fru-tue:bay}, we do not attempt to identify exact zeros in a covariance matrix. Rather, we aim to find a parsimonious factor representation of the underlying heteroskedastic data which may (or may not) imply covariances that are close to zero at times.}
In the real-world application, we evaluate our model against a wide range of alternative specifications via log predictive scores and minimum variance portfolio returns. Factor SV models with sufficiently many factors turn out to imply extremely competitive portfolios in relation to well-established methods which typically have been specifically tailored for such applications. Concerning density forecasts, we find that our approach outperforms all included competitors by a large margin. 

The remainder of this paper is structured as follows. In Section~\ref{sec:modspec}, the factor SV model is specified and the choice of prior distributions is discussed. Section~\ref{sec:estimation} treats statistical inference via MCMC methods and sheds light on computational aspects concerning out-of-sample density predictions for this model class. Extensive simulation studies are presented in Section~\ref{sec:simstud}, where the effect of the Normal-Gamma prior on correlation estimates is investigated in detail. In Section~\ref{sec:app}, the model is applied to 300 S\&P 500 members. Section~\ref{sec:concl} wraps up and points out possible directions for future research.




\section{Model Specification}
\label{sec:modspec}

 
Consider an $\dimy$-variate zero-mean return vector $\ym_t = \trans{(y_{1t}, \ldots,  y_{\dimy t})}$ for time $t = 1, \dots, T$ whose conditional distribution is Gaussian, i.e.
\[
 \ym_t|\Vary_t \sim \Normult{\dimy}{\bfz, \Vary_t}.
\]
\subsection{Factor SV Model}
To reduce dimensionality, factor SV models utilize a decomposition of the $\dimy \times \dimy$ covariance matrix $\Vary_t$ with $\dimy(\dimy+1)/2$ free elements into a factor loadings matrix $\facload$ of size $\dimy \times r$, an $r$-dimensional diagonal matrix $\Vm_t$ and an $\dimy$-dimensional diagonal matrix $\Vare_t$ in the following fashion:
\begin{equation}
\Vary_t = \facload \Vm_t \facload' + \Vare_t.
\label{vardeco}
\end{equation}
This reduces the number of free elements to $\dimy r + \dimy + r$. Because $r$ is typically chosen to be much smaller than $\dimy$, this specification constrains the parameter space substantially, thereby inducing parameter parsimony. For the paper at hand, $\facload$ is considered to be time invariant whereas the elements of both $\Vm_t$ and $\Vare_t$ are allowed to evolve over time through parametric stochastic volatility models, i.e.\ $\Vare_t = \Diag{\exp(\hsv_{1t}),\ldots,\exp(\hsv_{\dimy t})}$ and $\Vm_t = \Diag{\exp(\hsv_{\dimy+1,t}),\ldots,\exp(\hsv_{\dimy+r,t})}$ with 
\begin{eqnarray}  
 &&  \hsv_{it} \sim \Normal{\mupar_i +  \phipar_i (\hsv_{i,t-1}- \mupar_i),  \sigmapar_i^2}, \quad i = 1,\dots,m, \label{fac3newidi} \\
 &&  \hsv_{m+j,t} \sim \Normal{\phipar_{m+j} \hsv_{m+j,t-1},  \sigmapar_{m+j}^2}, \quad j = 1,\dots,r. \label{fac3newlat}
\end{eqnarray}
More specifically, $\Vare_t$ describes the idio\-syncratic (series-specific) variances while $\Vm_t$ contains the variances of underlying orthogonal factors  $\facm_t \sim \Normult{r}{\bfz, \Vm_t}$ that govern the contemporaneous dependence. The autoregressive process in (\ref{fac3newlat}) is assumed to have mean zero to identify the unconditional scaling of the factors.

This setup is commonly written in the following hierarchical form \citep[e.g.][]{chi-etal:ana}:
\begin{eqnarray*}  
\ym_t|\facload,\facm_t,\Vare_t \sim \Normult{\dimy}{\facload \facm_t,\Vare_t}, \quad \facm_t|\Vm_t \sim \Normult{\nfactrue}{\bfz,\Vm_t}, 
\end{eqnarray*}
where the distributions are assumed to be conditionally independent for all points in time. 
To make further exposition clearer, let $\ym = (\ym_1 \cdots \ym_T)$ denote the $m \times T$ matrix of all observations, $\fm=(\facm_1 \cdots \facm_T)$ the $r \times T$ matrix of all latent factors and  $\hsvm = (\hsvm_1'\cdots\hsvm_{m+r}')'$ the $(T+1)\times(m+r)$ matrix of all $m+r$ log-variance processes $\hsvm_i=(\hsv_{i0}, \hsv_{i1}, \ldots,\hsv_{iT})$, $i = 1,\dots,m+r$.
The vector $\allpara_i = (\mupar_i, \phipar_i, \sigmapar_i)'$ is referred to as the vector of parameters where $\mupar_i$ is the level, $\phipar_i$ the persistence, and $\sigmapar_i^2$ the innovation variance of $\hsvm_i$. To denote specific rows and columns of matrices, we use the ``dot'' notation, i.e.~$\bm{X}_{i\cdot}$ refers to the $i$th row and $\bm{X}_{\cdot j}$ to the $j$th column of $\bm{X}$.
The proportions of variances explained through the common factors for each component series, $C_{it} = 1-{\vare_{ii,t}}/{\vary_{ii,t}}$ for $i = 1,\dots,m,$ are referred to as the \emph{communalities}. Here, $\vare_{ii,t}$ and $\vary_{ii,t}$ denote the $i$th diagonal element of $\Vare_t$ and $\Vary_t$, respectively. As by construction $0 \leq \vare_{ii,t} \leq \vary_{ii,t}$, the communality for each component series and for all points in time lies between zero and one. The joint (overall) communality $C_t = m^{-1}\sum_{i=1}^m C_{it}$ is simply defined as the arithmetic mean over all series.

Three comments are in order.
First, the variance-covariance decomposition in (\ref{vardeco}) can be rewritten as $\Vary_t = \facload_t \facload'_t + \Vare_t$ with $\facload_t := \facload \Vm_t^{1/2}$. An essential assumption within the factor framework is that both $\Vm_t$ as well as $\Vare_t$ are diagonal matrices. This implies that the factor loadings $\facload_t$ are dynamic but can only vary column-wise over time. Consequently, the time-variability of $\Vary_t$'s off-diagonal elements are cross-sectionally restricted while its diagonal elements are allowed to move independently across series. Hence, the ``strength'' of a factor, i.e.\ its cross-sectional explanatory power, varies jointly for all series loading on it. Consequently, it is likely that more factors are needed to properly explain the co-volatility dynamics of a multivariate time series than in models which allow for completely unrestricted time-varying factor loadings \citep{lop-car:fac}, correlated factors \citep[$\Vm_t$ not diagonal, see][]{zho-etal:bay}, or approximate factor models \citep[$\Vare_t$ not diagonal, see][]{bai-ng:det}. Our specification, however, is less prone to overfitting and has the significant advantage of vastly simplified computations.

Second, identifying loadings for latent factor models is a long-standing issue that goes back to at least \cite{and-rub:sta} who discuss identification of factor loadings. Even though this problem is alleviated somewhat when factors are allowed to exhibit conditional heteroskedasticity \citep{sen-fio:ide, rig:ide}, most authors have chosen an upper triangular constraint of the loadings matrix with unit diagonal elements, thereby introducing dependence on the ordering of the data \citep[see][]{fru-lop:par}. However, when estimation of the actual factor loadings is not the primary concern (but rather a means to estimate and predict the covariance structure), this issue is less striking because a unique identification of the loadings matrix is not necessary.\footnote{The conditional covariance matrix $\Vary_t = \facload \Vm_t \facload' + \Vare_t$ involves a rotation-invariant transformation of $\facload$.} This allows leaving the factor loadings matrix completely unrestricted, thus rendering the method invariant with respect to the ordering of the series.

Third, note that even though the joint distribution of the data is conditionally Gaussian, its stationary distribution has thicker tails. Nevertheless, generalizations of the univariate SV model to cater for even more leptokurtic distributions \citep[e.g.][]{lie-jun:sto} or asymmetry \citep[e.g.][]{yu:on} can straightforwardly be incorporated in the current framework.
All of these extensions, however, tend to increase both sampling inefficiency as well as running time considerably and could thus preclude inference in very high dimensions.

\subsection{Prior Distributions}

The usual prior for each (unrestricted) element of the factor loadings matrix is a zero-mean Gaussian distribution, i.e.\ $\load_{ij} \sim \Normal{0, \tau^2_{ij}}$ independently for each $i$ and $j$, where $\tau^2_{ij} \equiv \tau^2$ is a constant specified a priori \citep[e.g.][]{pit-she:tim,agu-wes:bay,chi-etal:ana,ish-omo:por,kas-etal:eff}. To achieve more shrinkage, we model this variance hierarchically by placing a hyperprior on $\tau^2_{ij}$. 
This approach is related to \cite{bha-dun:spa,pat-etal:pos} who investigate a similar class of priors for homoskedastic factor models.
More specifically, let 
\begin{equation}
 \label{ngprior}
\load_{ij}|\tau_{ij}^2 \sim \Normal{0, \tau_{ij}^2}, \quad \tau_{ij}^2|\lambda_i^2 \sim \Gammad{a_i, a_i\lambda_i^2/2}, \quad \lambda_i^2 \sim \Gammad{c_i, d_i}.
\end{equation}
Intuitively, each prior variance $\tau_{ij}^2$ provides element-wise shrinkage governed independently for each row by $\lambda_i^2$.
Integrating out $\tau^2_{ij}$ yields a density for $\load_{ij}|\lambda_i^2$ of the form
$
p(\load_{ij}|\lambda_i^2) \propto |\load_{ij}|^{a_i-1/2}K_{a_i-1/2}(\sqrt a_i \lambda_i|\load_{ij}|),
$
where $K$ is the modified Bessel function of the second kind. This implies that the conditional variance of $\load_{ij}|\lambda_i^2$ is $2/\lambda_i^2$ and the excess kurtosis of $\load_{ij}$ is $3/a_i$.
The hyperparameters $a_i$, $c_i$, and $d_i$ are fixed a priori, whereas $a_i$ in particular plays a crucial role for the amount of shrinkage this prior implies. Choosing $a_i$ small enforces strong shrinkage towards zero, while choosing $a_i$ large imposes little shrinkage. For more elaborate discussions on Bayesian shrinkage in general and the effect of $a_i$ specifically, see \cite{gri-bro:inf} and \cite{pol-sco:shr}. Note that the Bayesian Lasso prior \citep{par-cas:bay} arises as a special case when $a_i = 1$.

One can see prior (\ref{ngprior}) as \emph{row-wise shrinkage} with \emph{element-wise adaption} in the sense that all variances in row $i$ can be thought of as ``random effects'' from the same underlying distribution. In other words, each series has high and a priori independent mass not to load on any factors and thus can be thought of as \emph{series-specific} shrinkage. For further aspects on introducing hierarchical prior structure via the Normal-Gamma distribution, see \cite{gri-bro:hie, hub-fel:ada}.
Analogously, it turns out to be fruitful to also consider \emph{column-wise shrinkage} with element-wise adaption, i.e.
\begin{equation*}
\load_{ij}|\tau_{ij}^2 \sim \Normal{0, \tau_{ij}^2}, \quad \tau_{ij}^2|\lambda_j^2 \sim \Gammad{a_j, a_j\lambda_j^2/2}, \quad \lambda_j^2 \sim \Gammad{c_j, d_j}.
\end{equation*}
This means that each factor has high and a priori independent mass not to be loaded on by any series and thus can be thought of as \emph{factor-specific} shrinkage.

Concerning the univariate SV priors, we follow \cite{kas-fru:anc}. For the $m$ idiosyncratic and $r$ factor volatilities, the initial states $\hsv_{i0}$ are distributed according to the stationary distributions of the AR($1$) processes (\ref{fac3newidi}) and (\ref{fac3newlat}), respectively.
Furthermore, $p(\mu_i,\phi_i,\sigma_i)$ = $p(\mu_i)p(\phi_i)p(\sigma_i)$, where the level $\mupar_i \in \mathbb{R}$ is equipped with the usual Gaussian prior $\mu_i \sim \Normal{b_{\mu}, B_{\mu}}$, the persistence parameter $\phipar_i \in (-1,1)$ is implied by $(\phipar_i+1)/2 \sim \Betad{a_0, b_0}$ and the volatility of volatility parameter $\sigmapar_i \in \mathbb{R}^+$ is chosen according to $\sigma_i^2 \sim B_\sigma \chi^2_1=\Gammad{1/2,1/(2B_\sigma)}$.

\section{Statistical Inference}
\label{sec:estimation}
There are a number of methods to estimate factor SV models such as quasi-maximum likelihood \citep[e.g.][]{har-etal:mul}, simulated maximum likelihood \citep[e.g.][]{lie-ric:cla,jun-koo:msv}, and Bayesian MCMC simulation \citep[e.g.][]{pit-she:tim, agu-wes:bay, chi-etal:ana, han:ass}. 
For high dimensional problems of this kind, Bayesian MCMC estimation proves to be a very efficient estimation method because it allows simulating from the high dimensional joint posterior by drawing from lower dimensional conditional posteriors.

\subsection{MCMC Estimation}
One substantial advantage of MCMC methods over other ways of learning about the posterior distribution is that it constitutes a modular approach due to the conditional nature of the sampling steps. Consequently, conditionally on the matrix of variances $\bm{\tau} = (\tau_{ij})_{1\leq i\leq m;\, 1\leq j\leq r}$, we can adapt the sampling steps of \cite{kas-etal:eff}. For obtaining draws for $\bm{\tau}$, we follow \cite{gri-bro:inf}. The MCMC sampling steps for the factor SV model are:
 \begin{itemize}
  \item[1.] For factors and idiosyncratic variances, obtain $m$ conditionally independent draws of the idiosyncratic log-volatilities from $\hsvm_{i}|\ym_{i\cdot},\facload_{i\cdot},\facm, \mu_i, \phi_i, \sigma_i$ and their parameters from $\mu_i, \phi_i, \sigma_i|\ym_{i\cdot},\facload_{i\cdot},\facm,\hsvm_{i}$ for $i = 1,\dots,m$. Similarly, perform $r$ updates for the factor log-volatilities from $\hsvm_{m+j}|\facm_{m+j,\cdot},\phi_{m+j}, \sigma_{m+j}$ and their parameters from $\phi_{m+j}, \sigma_{m+j}|\facm_{m+j,\cdot},\hsvm_{m+j}$ for $j = 1,\dots,r$. This amounts to $m+r$ univariate SV updates.\footnote{There is a vast body of literature on efficiently sampling univariate SV models. For the paper at hand, we use
    \proglang{R} package \pkg{stochvol} \citep{kas:dea}.
   }

  \item[2a.] Row-wise shrinkage only: For $i=1,\dots,m$, sample from 
   \[
    \lambda_i^2|\boldsymbol{\tau}_{i\cdot} \sim \Gammad{c_i+a_i \tilde r, d_i + \frac{a_i}{2}\sum_{j=1}^{\tilde r}\tau_{ij}^2},
   \]
   where $\tilde r = \min(i,r)$ if the loadings matrix is restricted to have zeros above the diagonal and $\tilde r = r$ in the case of an unrestricted loadings matrix. For $i=1,\dots,m$ and $j=1,\dots,\tilde r$, draw from $\tau_{ij}^2|\lambda_i,\load_{ij}
\sim
\text{GIG}(a_i-\frac{1}{2}, a_i\lambda_i^{2}, \load_{ij}^2)$.\footnote{\label{gignote}The Generalized Inverse Gaussian distribution $\text{GIG}(m,k,l)$ has a density proportional to $x^{m-1}\exp\left\{-\frac{1}{2}(kx+l/x)\right\}$. To draw from this distribution, we use the
algorithm described in \cite{hoe-ley:gen}
which is implemented in the \proglang{R} package \pkg{GIGrvg} \citep{r:gig}.
}

 \item[2b.] Column-wise shrinkage only: For $j=1,\dots,r$, sample from
  \[
  \lambda_j^2|\boldsymbol{\tau}_{\cdot j} \sim \Gammad{c_j+a_j (m-\tilde j+1), d_j + \frac{a_j}{2}\sum_{i=\tilde j}^m\tau_{ij}^2},
 \]
 where $\tilde j = j$ if the loadings matrix is restricted to have zeros above the diagonal and $\tilde j = 1$ otherwise.
 For $j=1,\dots,r$ and $i=\tilde j,\dots,r$, draw from $\tau_{ij}^2|\lambda_j,\load_{ij} \sim \text{GIG}(a_j-\frac{1}{2}, a_j\lambda_j^{2}, \load_{ij}^2)$.\textsuperscript{\ref{gignote}}

\item[3.] Letting $\boldsymbol{\Psi}_i = \text{diag}\left(\tau_{i1}^{-2}, \tau_{i2}^{-2}, \dots, \tau_{i\tilde r}^{-2}\right)$, draw $\facload_{i\cdot}'|\facm,\ym_{i\cdot},\hsvm_i,\boldsymbol{\Psi}_i, \sim \Normult{\tilde r}{\bm{b}_{iT}, \bm{B}_{iT}}$ with $\bm{B}_{iT}=(\bm{X}_i'\bm{X}_i + \boldsymbol{\Psi}_i)^{-1}$ and $\bm{b}_{iT}=\bm{B}_{iT}\bm{X}_i'\tilde \ym_{i\cdot}$. Hereby, $\tilde \ym_{i\cdot}=(y_{i1}\e^{-h_{i1}/2},\dots,y_{iT}\e^{-h_{iT}/2})'$ denotes the $i$th normalized observation vector and
\[
\bm{X}_i=
\begin{bmatrix}
\fac_{11}\e^{-h_{i1}/2}& \cdots & \fac_{\tilde r 1}\e^{-h_{i1}/2} \\
\vdots & & \vdots \\
\fac_{1T}\e^{-h_{iT}/2}& \cdots & \fac_{\tilde r T}\e^{-h_{iT}/2}
\end{bmatrix}
\]
is the $T\times \tilde r$ design matrix. This constitutes a standard Bayesian regression update.

\item[3*.] When inference on the factor loadings matrix is sought, optionally redraw $\facload$ using \emph{deep interweaving} \citep{kas-etal:eff} to speed up mixing. This step is of less importance if one is interested in the (predictive) covariance matrix only.
\item[4.] Draw the factors from $\facm_{t}|\facload,\ym_{t},\hsvm_{t} \sim \Normult{r}{\bm{b}_{mt}, \bm{B}_{mt}}$ with $\bm{B}_{mt}^{-1}=\bm{X}_t'\bm{X}_t + \Vm_t^{-1}$ and $\bm{b}_{mt}=\bm{B}_{mt}\bm{X}_t'\tilde \ym_t$. Hereby, $\tilde \ym_{t}=(y_{1t}\e^{-h_{1t}/2},\dots,y_{mt}\e^{-h_{mt}/2})'$ denotes the normalized observation vector at time $t$ and
\[
\bm{X}_t=\begin{bmatrix}
\load_{11}\e^{-h_{1t}/2} & \cdots & \load_{1r}\e^{-h_{1t}/2}\\
\vdots&&\vdots\\
\load_{m1}\e^{-h_{mt}/2} & \cdots & \load_{mr}\e^{-h_{mt}/2}\\
\end{bmatrix}
\]
is the $m\times r$ design matrix. This constitutes a standard Bayesian regression update.
\end{itemize}

\begin{table}[t]
\centering
\caption{Empirically obtained runtime per MCMC iteration on a single i5-5300U CPU (2.30GHz) core running Xubuntu Linux 16.04 using \proglang{R} 3.3.3 linked against the default linear algebra packages as well as Intel MKL (single thread). Measured in milliseconds for $m \in \{10, 100, 500\}$ dimensional time series of length $T = 1000$.}
\label{tab:runtime}
\begin{tabular}{lrrrrrr}
 \hline
  & $r=0$ & $r=1$ & $r=5$ & $r=10$ & $r=20$ & $r=50$\\
  \hline
plain \proglang{R}, $m=10  $ & 4 & 5   &  8   &   14  &     &  \\
MKL, $m=10 $ & 5 & 6  &  10  &  15 &  &  \\
plain \proglang{R}, $m=100 $ & 43 & 46  &  56  & 74  & 131 & 451 \\
MKL, $m=100 $ & 45 & 49  &  57  & 69  & 90 & 185 \\
plain \proglang{R}, $m=500 $ & 222 & 240 & 279  & 361 & 600 & 1993 \\
MKL, $m=500 $ & 225 & 240 & 269 & 310 & 389 & 693 \\

\hline
\end{tabular}
\end{table}

The above sampling steps are implemented in an efficient way within the \proglang{R} package \pkg{factorstochvol} \citep{r:fac}.
Table~\ref{tab:runtime} displays the empirical run time in milliseconds per MCMC iteration. Note that using more efficient linear algebra routines such as Intel MKL leads to substantial speed gains only for models with many factors. To a certain extent, computation can further be sped up by computing the individual steps of the posterior sampler in parallel. In practice, however, doing so is only useful in shared memory environments (e.g.\ through multithreading/multiprocessing) as the increased communication overhead in distributed memory environments easily outweighs the speed gains.

\subsection{Prediction}

Given draws of the joint posterior distribution of parameters and latent variables, it is in principle straightforward to predict future covariances and consequently also future observations. This gives rise to the predictive density \citep{gew-ami:com}, defined as
\begin{eqnarray}
 \label{preddens}
p(\yb_{t+1}|\yb^o_{[1:t]}) = \int\limits_{\bm K} \!
 p(\yb_{t+1}|\yb^o_{[1:t]}, \unobs) \times
 p(\unobs|\yb^o_{[1:t]}) 
 \, \dif \unobs,
\end{eqnarray}
where $\unobs$ denotes the vector of all unobservables, i.e.\ parameters and latent variables.
The superscript $o$ in $\yb^o_{[1:t]}$ denotes \emph{ex post} realizations (observations) for the set of points in time $\{1,\dots ,t\}$ of the \emph{ex ante} random values $\yb_{[1:t]} = (\yb_1 \cdots \yb_t)$. The integration space ${\bm K}$ simply stands for the space of the possible values for $\unobs$. Because (\ref{preddens}) is the integral of the likelihood function where the values of $\unobs$ are weighted according to their posterior distribution, it can be seen as the forecast density for an unknown value $\yb_{t+1}$ after accounting for the uncertainty about $\unobs$, given the history $\yb^o_{[1:t]}$.

As with most quantities of interest in Bayesian analysis, computing the predictive density can be challenging because it constitutes an extremely high-dimensional integral which cannot be solved analytically. However, it may be approximated at a given ``future'' point $\yb^f$ through Monte Carlo integration,
\begin{equation}
 \label{preddensMC}
 p(\yb^f|\yb^o_{[1:t]}) \approx \frac{1}{\M}\sum_{\m=1}^\M p(\yb^f|\yb^o_{[1:t]}, \unobs^{(\m)}_{[1:t]}),
\end{equation}
where $\unobs^{(\m)}_{[1:t]}$ denotes the $\m${th} draw from the posterior distribution up to time $t$. If (\ref{preddensMC}) is evaluated at $\yb^f=\yb_{t+1}^o$, it is commonly referred to as the (one-step-ahead) \emph{predictive likelihood} at time $t+1$, denoted $\PL_{t+1}$.
Also, draws from (\ref{preddens}) can straightforwardly be obtained by generating values $\yb_{t+1}^{(\m)}$ from the distribution given through the (in our case multivariate Gaussian) density $p(\yb_{t+1}|\yb^o_{[1:t]}, \unobs^{(\m)}_{[1:t]})$.

For the model at hand, two ways of evaluating the predictive likelihood particularly stand out. First, one could average over $\m=1,\dots, \M$ densities of $$\Normult{m}{\facload^{(\m)}_{[1:t]}\facm^{(\m)}_{t+1,[1:t]}, \Vare^{(\m)}_{t+1,[1:t]}},$$
    evaluated at $\yb^o_{t+1}$, where the subscript $t+1$ denotes the corresponding one-step ahead predictive draws and
    $\Vare^{(\m)}_{t+1,[1:t]} = \Diag{\exp{\hsv_{1,t+1,[1:t]}^{(\m)}}, \dots, \exp{\hsv_{\dimy,t+1,[1:t]}^{(\m)}}}$. Note that because $\Vare^{(\m)}_{t+1,[1:t]}$ is by construction diagonal, this method only requires univariate Gaussian evaluations and is thus computationally efficient. Nevertheless, because evaluation is done conditionally on realized values of $\facm_{t+1,[1:t]}$, it is extremely unstable in many dimensions. Moreover, since the numerical inaccuracy increases with an increasing number of factors $r$, this approach can lead to systematic undervaluation of $\PL_{t+1}$ for larger $r$. Thus, in what follows, we recommended an alternative approach.

To obtain $\PL_{t+1}$, we suggest to average over $\m=1,\dots, \M$ densities of $$\Normult{m}{\bfz, \facload^{(\m)}_{[1:t]} \Vm^{(\m)}_{t+1,[1:t]} (\facload^{(\m)}_{[1:t]})' + \Vare^{(\m)}_{t+1,[1:t]}},$$
evaluated at $\yb^o_{t+1}$, where $\Vm^{(\m)}_{t+1,[1:t]} = \Diag{\exp{\hsv_{\dimy+1,t+1,[1:t]}^{(\m)}}, \dots, \exp{\hsv_{\dimy+r,t+1,[1:t]}^{(\m)}}}$. This form of the predictive likelihood is obtained by analytically performing integration in (\ref{preddens}) with respect to $\facm_{t+1,[1:t]}$. Consequently, it is numerically more stable, irrespectively of the number of factors $r$. However, it requires a full $\dimy$-variate Gaussian density evaluation for each $k$ and is thus computationally much more expensive. To a certain extent, the computational burden can be mitigated by using the Woodbury matrix identity, $\Vary_t^{-1}
    = \Vare_t^{-1} - \Vare_t^{-1}\facload \left(\Vm_t^{-1}+\facload'\Vare_t^{-1}\facload \right)^{-1} \facload'\Vare_t^{-1}$, 
along with the matrix determinant lemma, $\det(\Vary_t)
= \det(\Vm_t^{-1} + \facload'\Vare_t^{-1}\facload) \det(\Vm_t)\det(\Vare_t)$. This substantially speeds up the repetitive evaluation of the multivariate Gaussian distribution if $r \ll m$.


We apply these results for comparing competing models $A$ and $B$ between time points $t_1$ and $t_2$ and consider cumulative log predictive Bayes factors defined through $\log \BF_{t_1,t_2}(A, B) = \sum_{t=t_1+1}^{t_2} \log \PL_t(A) - \log \PL_t(B)$, where $\PL_t(A)$ and $\PL_t(B)$ denote the predictive likelihood of model $A$ and $B$ at time $t$, respectively. When the cumulative log predictive Bayes factor is greater than 0 at a given point in time, there is evidence in favor of model $A$, and vice versa. Thereby, data up to time $t_1$ is regarded as prior information and out-of-sample evaluation starts at time $t_1+1$.

\section{Simulation Studies}
\label{sec:simstud}
The aim of this section is to apply the model to a simulated data set in order to illustrate the shrinkage properties of the Normal-Gamma prior for the factor loadings matrix elements. For this purpose, we first illustrate several scenarios on a single ten dimensional data set. Second, we investigate the performance of our model in a full Monte Carlo simulation based on $100$ simulated data sets. Third, and finally, we investigate to what extend these results carry over to higher dimensions.

In what follows, we compare five specific prior settings. Setting~1 refers to the usual standard Gaussian prior with variance $\tau_{ij}^2 \equiv \tau^2 = 1$ and constitutes the benchmark. Setting~2 is the row-wise Bayesian Lasso where $a_i = 1$ for all $i$. Setting~3 is the column-wise Bayesian Lasso where $a_j=1$ for all $j$. Setting~4 is the Normal-Gamma prior with row-wise shrinkage where $a_i = 0.1$ for all $i$. Setting~5 is the Normal-Gamma prior with column-wise shrinkage where $a_j = 0.1$ for all $j$. Throughout this section, prior hyperparameters are chosen as follows: $b_\mu = 0$, $B_\mu = 1000$, $B_\sigma = 1$. The prior hyperparameters for the persistence of the latent log variances are fixed at $a_0 = 10$, $b_0 = 2.5$ for the idiosyncratic volatilities and $a_0 = 2.5$, $b_0 = 2.5$ for the factor volatilities; note that the parameters of the superfluous factor are only identified through the prior. The shrinkage hyperparameters are set as in \cite{bel-etal:hie}, i.e.\ $c_i=c_j=d_i=d_j=0.001$ for all applicable $i$ and $j$. For each setting, the algorithm is run for $110\,000$ iterations of which the first $10\,000$ draws are discarded as burn-in.

\subsection{The Shrinkage Prior Effect: An Illustration}

\begin{figure}[p]
\includegraphics[width = \textwidth]{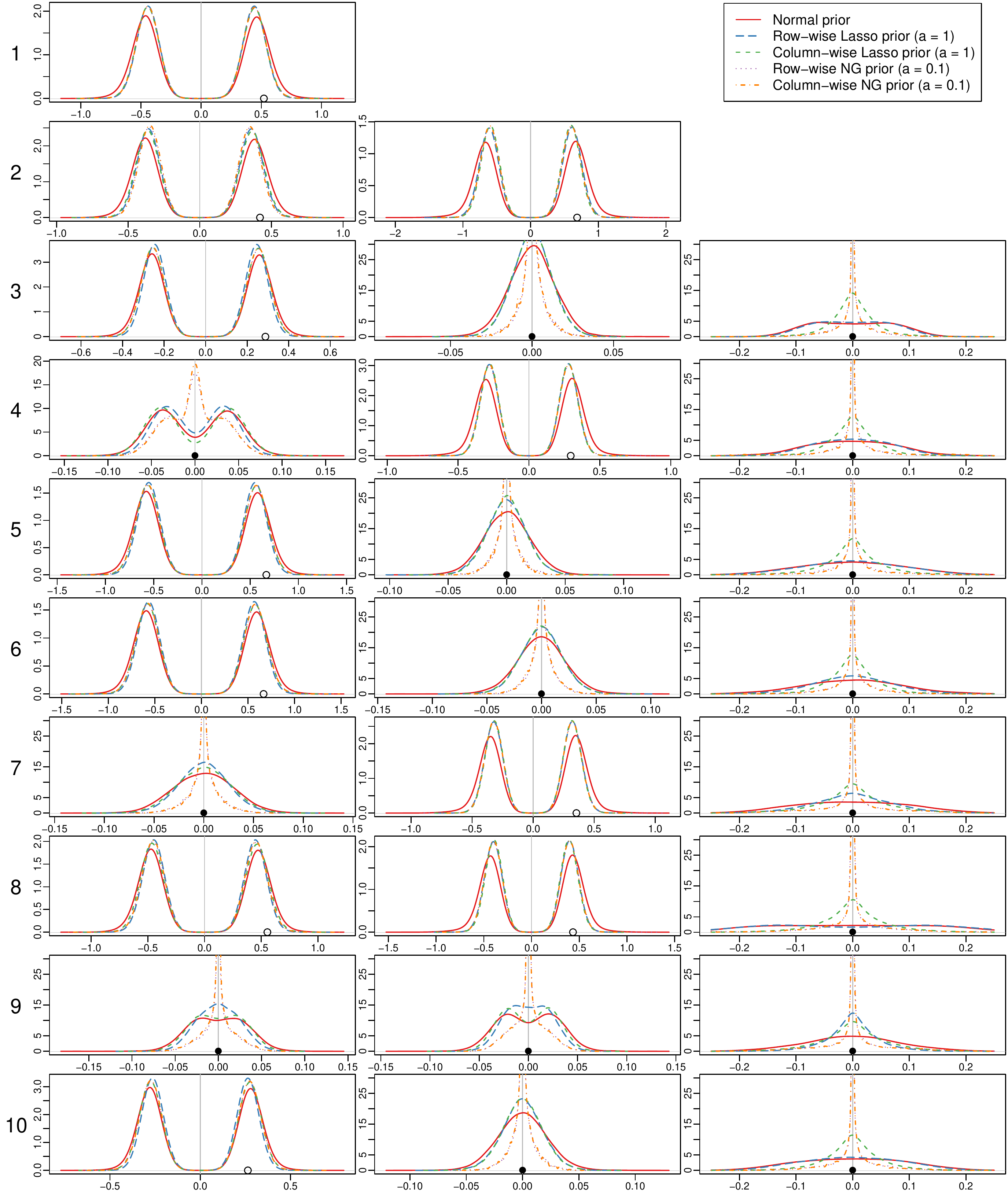}
\caption{Kernel density estimates of posterior factor loadings under different priors. The standard Gaussian prior (setting 1) in red solid strokes, the row-wise Lasso prior (setting 2) in blue long-dashed strokes, the column-wise Lasso prior (setting 3) in green short-dashed strokes, the row-wise Normal-Gamma prior (setting 4) in purple dotted strokes, the column-wise Normal-Gamma prior (setting 5) in orange dashed-dotted strokes. The vertical axis is capped at 30.}
\label{postload}
\end{figure}

To investigate the effects of different priors on the posteriors of interest, we simulate a single data set from a two factor model for $m=10$ time series of length $T=1000$. For estimation, an overfitting model with three latent factors is employed. The nonzero parameter values used for simulation are picked randomly and are indicated as black circles in Figure~\ref{postload}; some loadings are set to zero, indicated by black dots. We set $\load_{ij}$ to zero if $j>i$ for simulation and estimation.

Figure~\ref{postload} shows smoothed kernel density estimates of posterior loadings under the different prior assumptions. The signs of the loadings have not been identified so that a multimodal posterior distribution hints at a ``significant'' loading whereas a unimodal posterior hints at a zero loading, see also \cite{fru-wag:sto}.
It stands out that only very little shrinkage is induced by the standard Gaussian prior. The other priors, however, impose considerably tighter posteriors. For the nonzero loadings on factor one, e.g., the row-wise Bayesian Lasso exhibits the strongest degree of shrinkage. Little difference between the various shrinkage priors can be spotted for the nonzero loadings on factor two.

Turning towards the zero loadings, the strongest shrinkage is introduced by both variants of the Normal-Gamma prior, followed by the different variants of the Bayesian Lasso and the standard Gaussian prior. This is particularly striking for the loadings on the superfluous third factor. The difference between row- and column-wise shrinkage for the Lasso variants can most clearly be seen in row 9 and column 3, respectively. The row-wise Lasso captures the ``zero-row'' 9 better, while the column-wise Lasso captures the ``zero-column'' 3 better. Because of the increased element-wise shrinkage of the Normal-Gamma prior, the difference between the row-wise and the column-wise variant are minimal. 

\begin{figure}[t]
\includegraphics[width = \textwidth]{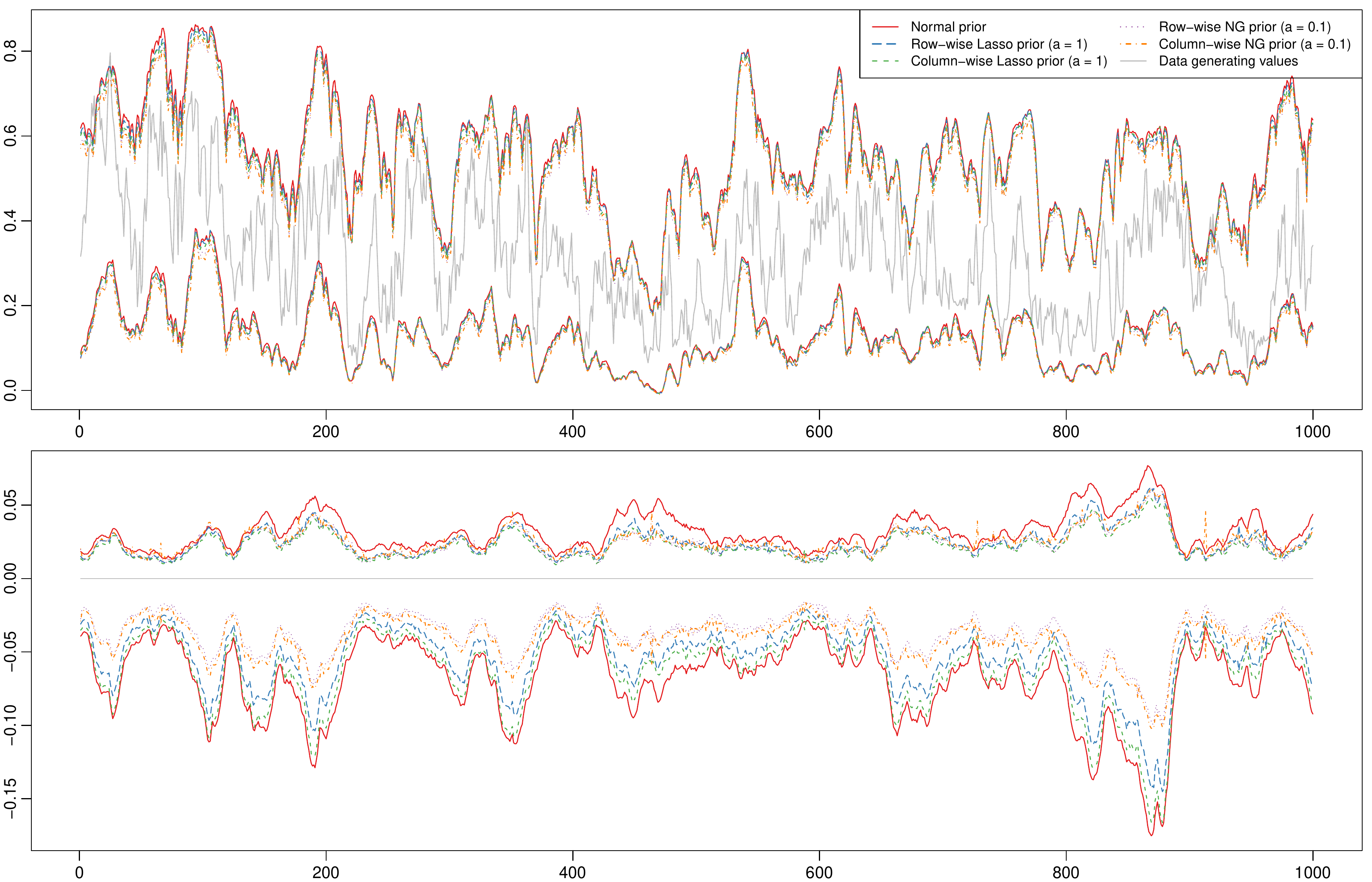}
\caption{``True'' (gray, solid) and estimated posterior correlations between series 1 and series 2 (top) as well as series 9 and series 10 (bottom). To illustrate estimation uncertainty, posterior means plus/minus 2 times posterior standard deviations are displayed.
}
\label{postcortime}
\end{figure}

In the context of covariance modeling, however, factor loadings can be viewed upon as a mere means to parsimony, not the actual quantity of interest.
Thus, Figure~\ref{postcortime} displays selected time-varying correlations. The top panel shows a posterior interval estimate (mean plus/minus two standard deviations) for the correlation of series 1 and series 2 (which is nonzero) under all five prior settings; the bottom panel depicts the interval estimate for the correlation of series 9 and 10 (which is zero). While the relative differences between the settings in the nonzero correlation case are relatively small, the zero correlation case is picked up substantially better when shrinkage priors are used. Posterior means are closer to zero and the posterior credible intervals are tighter.

\include*{tables/predictive_sim_10/tables}

To conclude, we briefly examine predictive performance by investigating cumulative log predictive Bayes factors. Thereby, the first $1000$ points in time are treated as prior information, then 1-day- and 10-days-ahead predictive likelihoods are recursively evaluated until $t=1500$. Table~\ref{pred10} displays the sum of these values for the respective models in relation to the 2-factor model with the standard Gaussian prior. This way, numbers greater than zero can be interpreted as evidence in favor of the respective model.
Not very surprisingly, log Bayes factors are highest for the 2-factor model; within this class, models imposing stronger shrinkage perform slightly better, in particular when considering the longer 10-day horizon. Underfitting models predict very poorly both on the short and the longer run, while overfitting models appear almost en par with the baseline model when shrinkage priors are used. This suggests that shrinkage safeguards against overfitting, at least to a certain extent.

\subsection{Medium Dimensional Monte Carlo Study}

For a more comprehensive understanding of the shrinkage effect, the above study is repeated for $100$ different data sets where all latent variables are generated randomly for each realization. In Table~\ref{rmsessmall}, the medians of the respective relative RMSEs (root mean squared errors, averaged over time) between the true and the estimated pairwise correlations are depicted. The part above the diagonal represents the relative performance of the row-wise Lasso prior (setting 2) with respect to the baseline prior (setting 1), the part below the diagonal represents the relative performance of the row-wise Normal-Gamma prior (setting 4) with respect to the row-wise Lasso prior (setting 2). Clearly, gains are highest for series 9 which is by construction completely uncorrelated to the other series.
Additionally, geometric averages of these performance indicators are displayed in the first row (setting 2 vs.\ baseline) and in the last row (setting 4 vs.\ baseline). They can be seen as the average relative performance of one specific series' correlation estimates with all other series.

\include*{tables/10/rmses_per_series_rel_small}

To illustrate the fact that extreme choices of $c_i$ and $d_i$ are crucial for the shrinkage effect of the Bayesian Lasso, Table~\ref{rmseslarge} displays relative RMSEs for moderate hyperparameter choices $c_i = d_i = 1$. Note that the performance of the Bayesian Lasso deteriorates substantially while performance of the Normal-Gamma prior is relatively robust with regard to these choices. This indicates that the shrinkage effect of the Bayesian Lasso is strongly dependent on the particular choice of these hyperparameters (governing row-wise shrinkage) while the Normal-Gamma can adapt better through increased element-wise shrinkage.

\include*{tables/10/rmses_per_series_rel_large}

An overall comparison of the errors under different priors is provided in Table~\ref{rmsesoverall} which lists RMSEs and MAEs for all prior settings, averaged over the non-trivial correlation matrix entries as well as time. Note again that results under the Lasso prior are sensitive to the particular choices of the global shrinkage hyperparameters as well as the choice of row- or column-wise shrinkage, which is hardly the case for the Norma-Gamma prior.
Interestingly, the performance gains achieved through shrinkage prior usage are higher when absolute errors are considered. This is coherent with the extremely high kurtosis of Normal-Gamma-type priors which, while placing most mass around zero, allow for large values.

\include*{tables/10/rmses_per_series_overall}

\subsection{High Dimensional Monte Carlo Study}

The findings are similar if dimensionality is increased; in analogy to above, we report overall RMSEs and MAEs for $495\,000$ pairwise correlations, resulting from $m=100$ component series at $T=1000$ points in time. The factor loadings for the $r = 10$ factors are again randomly sampled with 43.8\% of the loadings being equal to zero, resulting in about 2.6\% of the pairwise correlations being zero. Using this setting, $100$ data sets are generated; for each of these, a separate (overfitting) factor SV model using $r = 11$ factors without any prior restrictions on the factor loadings matrix is fit. The error measures are computed and aggregated. Table~\ref{rmsesoverall100} reports the medians thereof. In this setting, the shrinkage priors outperform the standard Gaussian prior by a relatively large margin; the effect of the specific choice of the global shrinkage hyperparameters is less pronounced.

\include*{tables/100/rmses_per_series_overall_cor}

\section{Application to S\&P 500 Data}
\label{sec:app}

In this section we apply the SV factor model to stock prices listed in the Standard \& Poor's 500 index. 
We only consider firms which have been continuously included in the index from November 1994 until December 2013, resulting in $m=300$ stock prices on $5001$ days, ranging from 11/1/1994 to 12/31/2013. The data was obtained from Bloomberg Terminal in January 2014. Instead of considering raw prices we investigate percentage log-returns which we demean a priori.


The presentation consists of two parts. First, we exemplify inference using a multivariate stochastic volatility model and discuss the outcome. Second, we perform out-of-sample predictive evaluation and compare different models.
To facilitate interpretation of the results discussed in this section, we consider the GICS\footnote{Global Industry Classification Standard, retrieved from \url{https://en.wikipedia.org/w/index.php?title=List_of_S\%26P_500_companies&oldid=589980759} on April 11, 2016.} classification into 10 sectors listed in Table~\ref{gics}.

\begin{table}[t]
\centering
\begin{tabular}{lr}
  \hline
GICS sector &  Members \\ 
  \hline
Consumer Discretionary &  45 \\ 
  Consumer Staples &  28 \\ 
  Energy &  23 \\ 
  Financials &  54 \\ 
  Health Care &  30 \\ 
  Industrials &  42 \\ 
  Information Technology &  27 \\ 
  Materials &  23 \\ 
  Telecommunications Services &   3 \\ 
  Utilities &  25 \\ 
   \hline
\end{tabular}
\caption{GICS sectors and the amount of members within the S\&P 500 data set.}
\label{gics}
\end{table}

\subsection{A Four-Factor Model for 300 S\&P 500 Members}

To keep graphical representation feasible, we only focus on the latest 2000 returns of our data set, i.e.\ 5/3/2006 to 12/31/2013. This time frame is chosen to include both the 2008 financial crisis as well as the period before and thereafter. Furthermore, we restrict our discussion to a four-factor model. This choice is somewhat arbitrary but allows for a direct comparison to a popular model based on four observed (Fama-French plus Momentum) factors. A comparison of predictive performance for varying number of factors is discussed in Section~\ref{prediction}; the Fama-French plus Momentum model is introduced in Section~\ref{sec:ff}.

\begin{figure}[t]
 \centering
 \includegraphics[width=\textwidth]{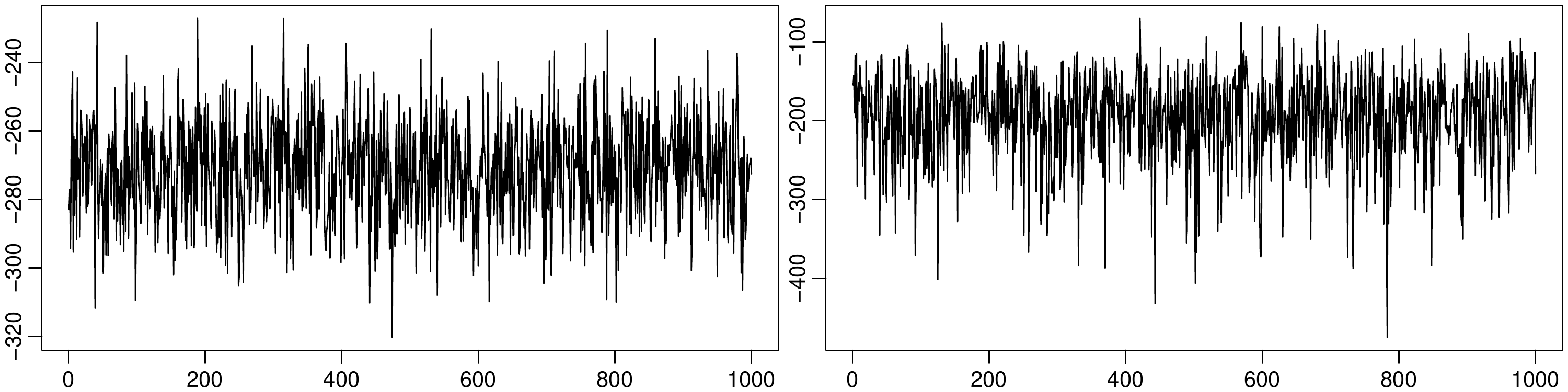}
 \caption{Trace plots for $1000$ draws after burn-in and thinning of the log determinant of the covariance (left panel) and the correlation matrix (right panel) for $t = T$.}
 \label{traces}
\end{figure}

We run our sampler employing the Normal-Gamma prior with row-wise shrinkage for $110\,000$ draws and discard the first $10\,000$ draws as burn-in.\footnote{To keep presentation at a reasonable length and because qualitative as well as quantitative results are very similar, we omit details about the Normal-Gamma prior with column-wise shrinkage.} Of the remaining $100\,000$ draws every $10$th draw is kept, resulting in $10\,000$ draws used for posterior inference. Hyperparameters are set as follows: $a_i \equiv a = 0.1$, $c_i \equiv c=1$, $d_i \equiv d = 1$, $b_\mu=0$, $B_\mu=100$, $a_0=20$, $b_0=1.5$, $B_\sigma=1, B_{m+j} = 1$ for $j = 1,\dots,r$. To prevent factor switching, we set all elements above the diagonal to zero. The leading series are chosen manually after a preliminary unidentified run such that series with high loadings on that particular factor (but low loadings on the other factors) become leaders. Note that this intervention (which introduces an order dependency) is only necessary for interpreting the factor loadings matrix but not for covariance estimation or prediction. Concerning MCMC convergence, we observe excellent mixing for both the covariance as well as the correlation matrix draws. To exemplify, trace plots of the first $1000$ draws after burn-in and thinning for posterior draws of the log determinant distribution of the covariance and correlation matrices at $t = T$ are displayed in Figure~\ref{traces}.

\begin{figure}[t]
 \centering
 \includegraphics[width=\textwidth]{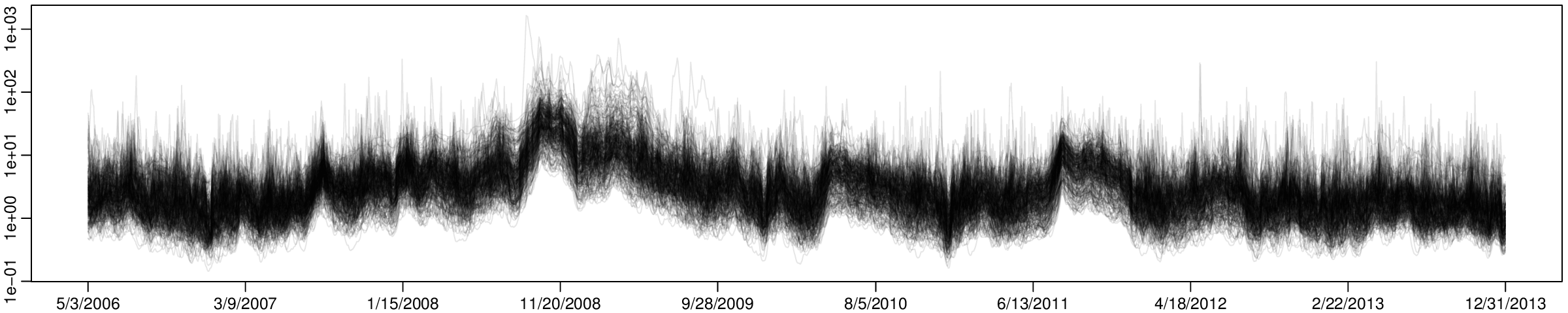}
 \includegraphics[width=\textwidth]{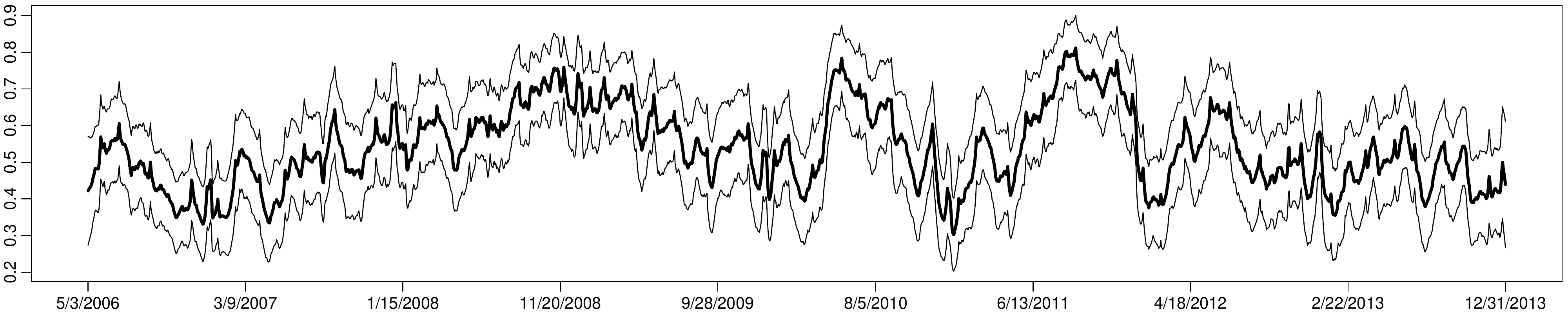}
 \caption{Top panel: 300 mean posterior variances, i.e.~$\mathbb{E}(\Diag{\Vary_t}|\ym)$ for $t = 1, \dots, T$ (logarithmic scale). Bottom panel: Posterior mean of the joint communalities $C_t$ (bold line) along with mean plus/minus two posterior standard deviations (light lines).}
 \label{varscommu}
\end{figure}

To illustrate the substantial degree of volatility co-movement, mean posterior variances are displayed in the top panel of Figure~\ref{varscommu}. This depiction resembles one where all series are modeled with independent univariate stochastic volatility models. Clear spikes can be spotted during the financial crisis in late 2008 but also in early 2010 and late 2011. This picture is mirrored (to a certain extent) in the bottom panel which displays the posterior distribution of the joint communality $C_{t}$.
In particular during the financial crisis, the first half of 2010 and late 2011 the joint communality reaches high values of 0.7 and more.

\begin{figure}[p]
 \centering
 \includegraphics[page=1, width=\textwidth, trim=0 0 0 28, clip]{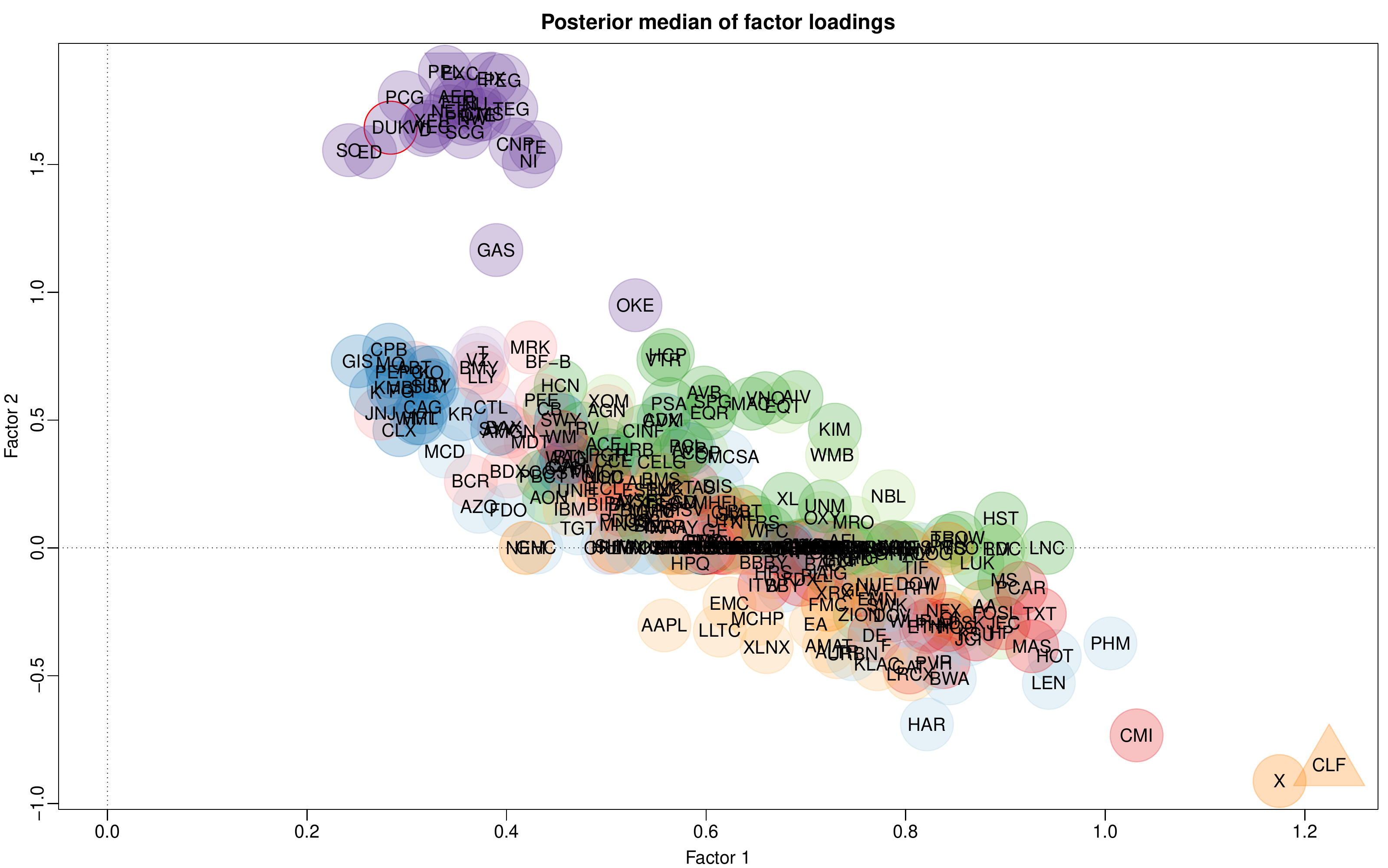}\\
 \vspace{1em}
 \includegraphics[page=2, width=\textwidth, trim=0 0 0 28, clip]{loadpoints.pdf}
 \caption{Median loadings on the first two factors (top) and the last two factors (bottom) of a $4$-factor model applied to $m=300$ demeaned stock price log-returns listed in the S\&P 500 index. Shading:
 Sectors according to the Global Industry Classification Standard.}
 \label{sp500facload}
\end{figure}

Median posterior factor loadings are visualized in Figure~\ref{sp500facload}. In the top panel it can be seen that all series significantly load on the first factor which consequently could be interpreted to represent the joint dynamics of the wider US equity market. Highly loading elements include United States Steel Corp.\ (X) and Cliffs Natural Resources Inc.\ (CLF), both of which belong to the sector \emph{Materials} and both of which have been dropped from the S\&P 500 index in 2014 due to market capitalization changes. Cummins Inc.\ (CMI, \emph{Industrials}) and PulteGroup, Inc.\ (PHM, \emph{Consumer Discretionary}) rank third and fourth. Companies in sectors \emph{Consumer Staples}, \emph{Utilities} and \emph{Health Care} tend to load comparably low on this factor.

Investigating the second factor, it stands out that due to the use of the Normal-Gamma prior a considerable amount of loadings are shrunk towards zero. Main drivers are all in the sector \emph{Utilities}. Also, companies in sectors \emph{Consumer Staples}, \emph{Health Care} and (to a certain extent) \emph{Financials} load positively here. Both the loadings on factor 3 as well as the loadings on factor 4 are substantially shrunk towards zero. Notable exceptions are \emph{Energy} and \emph{Materials} companies for factor 3 and \emph{Financials} for factor 4.

\begin{figure}[p]
 \centering
 \includegraphics[width=\textwidth]{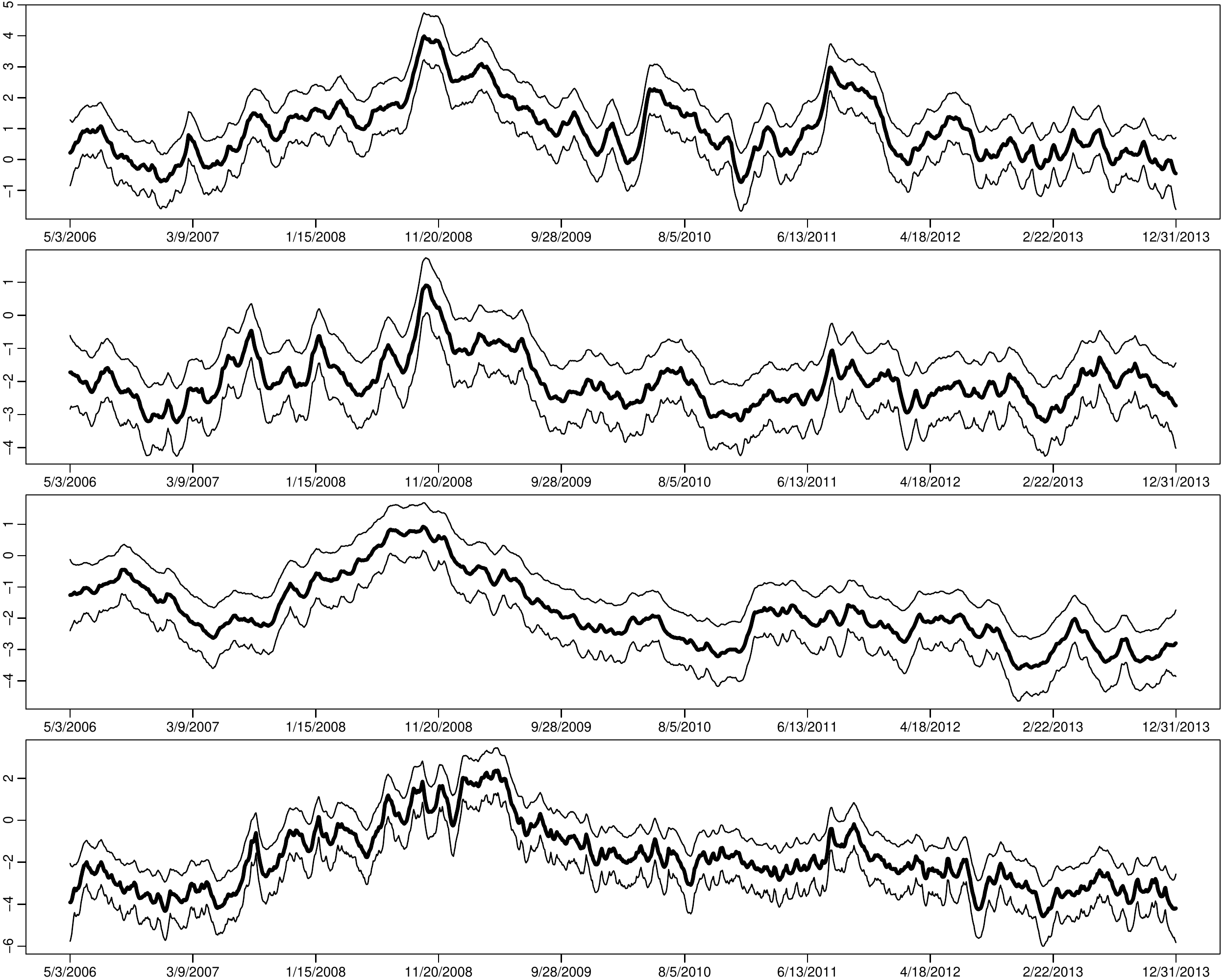}
 \caption{Latent factor log variances $\hsvm_{m+j,\cdot}$, $j=1,\dots,4$ (top to bottom). Bold line indicates the posterior mean; light lines indicate mean $\pm$ 2 standard deviations.}
 \label{faclogvars}
\end{figure}

The corresponding factor log variances are displayed in Figure~\ref{faclogvars}. Apart from featuring similar low- to medium frequency properties,
each process exhibits specific characteristics. First, notice the sharp increase of volatility in early 2010 which is mainly visible for the ``overall'' factor 1. The second factor (\emph{Utilities}) displays a pre-crisis volatility peak during early 2008. The third factor, driven by \emph{Energy} and \emph{Materials}, shows relatively smooth volatility behavior while the fourth factor, governed by the \emph{Financial}, exhibits a comparably ``nervous'' volatility evolution.

\begin{figure}[p]
\includegraphics[width=.327\textwidth]{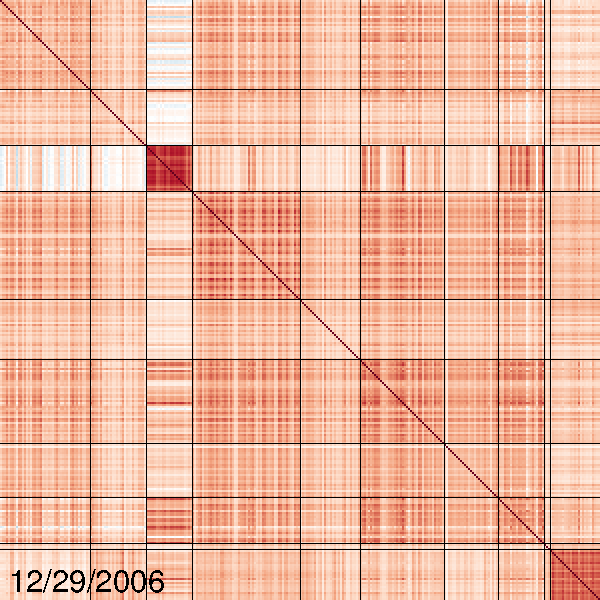}
\includegraphics[width=.327\textwidth]{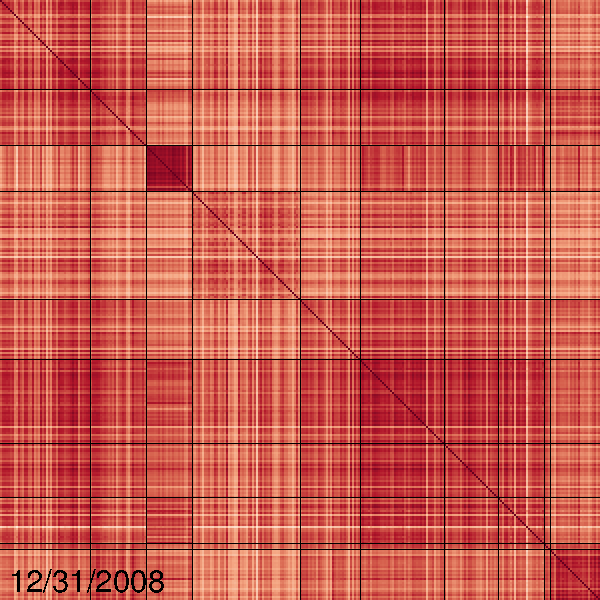}
\includegraphics[width=.327\textwidth]{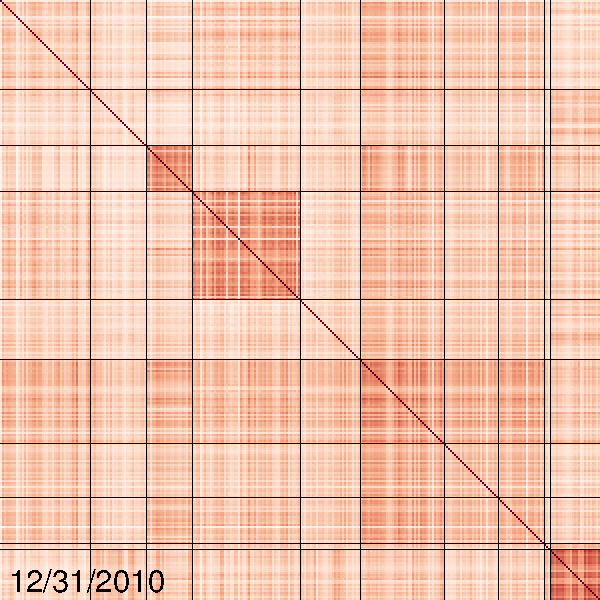}
\caption{Posterior mean of the time-varying correlation matrix $\mathbb{E}(\Vary_t|\ym)$, exemplified for $t\in \{173,696,1218\}$ which corresponds to the last trading day in 2006, 2008, 2010, respectively. The matrix has been rearranged to reflect the different GICS sectors in alphabetical order, cf.\ Table~\ref{gics}.}
\label{cormat}
\end{figure}

Finally, we show three examples of the posterior mean of the correlation matrix $\Vary_t$ in Figure~\ref{cormat}. The series are grouped according to the alphabetically ordered industry sectors (and simply sorted according to their ticker symbol therein).
An animation displaying the mean correlation matrix for all points in time is available at \url{https://vimeo.com/217021226}.

Considering the last trading day in 2006, highly correlated clusters appear within \emph{Energy} and \emph{Utilities}, to a certain extent also within \emph{Financials}, \emph{Industrials} and \emph{Materials}. Not very surprisingly, there exists only low correlation between companies in the sectors \emph{Consumer Discretionary/Staples} and \emph{Energy} but higher correlation between \emph{Energy}, \emph{Industrials} and \emph{Materials}.
Looking at the last trading day of 2008, the overall picture changes radically. Higher correlation can be spotted throughout, both within sectors but also between sectors. There are only few companies that show little and virtually no companies that show no correlation with others.
Another two years later, we again see a different overall picture. Lower correlations throughout become apparent with moderate correlations remaining within the sectors \emph{Energy}, \emph{Utilities}, and in particular \emph{Financials}.

\subsection{Predictive Likelihoods for Model Selection}
\label{prediction}

Even for univariate volatility models evaluating in- or out-of-sample fit is not straightforward because the quantity of interest (the conditional standard deviation) is not directly observable. While in lower dimensions this issue can be circumvented to a certain extent using intraday data and computing realized measures of volatility, the difficulty becomes more striking when the dimension increases. Thus, we focus on iteratively predicting the observation density out-of-sample which is then evaluated at the actually observed values. Because this approach involves re-estimating the model for each point in time, it is computationally costly but can be parallelized in a trivial fashion on multi-core computers.

For the S\&P 500 data set, we begin by using the first $3000$ data points (until 5/2/2006) to estimate the one-day-ahead predictive likelihood for day $3001$ as well as the ten-day-ahead predictive likelihood for day $3010$. In a separate estimation procedure, the first $1001$ data points (until 5/3/2006) are used to estimate the one-day-ahead predictive likelihood for day $3002$ and the corresponding ten-day-ahead predictive likelihood for day $3011$, etc. This procedure is repeated for $1990$ days until the end of the sample is reached.

We use a no-factor model as the baseline which corresponds to $300$ individual stochastic volatility models fitted to each component series separately. For each date, values greater than zero mean that the model outperforms the baseline model up to that point in time. Competitors of the no-factor SV model are $r$-factor SV models with $r = 1,\dots,20$ under the usual standard Gaussian prior and under the Normal-Gamma prior with $a_i \equiv 0.1$ and $c_i \equiv d_i \equiv 1$ employing row-wise-shrinkage. All other parameters are kept identical, i.e.~$b_\mu=0$, $B_\mu=100$, $a_0=20$, $b_0=2.5$ (idiosyncratic persistences), $a_0=2.5$, $b_0=2.5$ (factor persistences), $B_\sigma=0.1, B_{m+j} = 0.1$ for $j = 1,\dots,r$. Note that because the object of interest in this exercise does not require the factor loadings matrix to be identified, no a priori restrictions are placed on $\facload$. This alleviates the problem of arranging the data in any particular order before running the sampler. Other competing models are discussed in the following sections.

\begin{figure}[t]
\includegraphics[width=\textwidth]{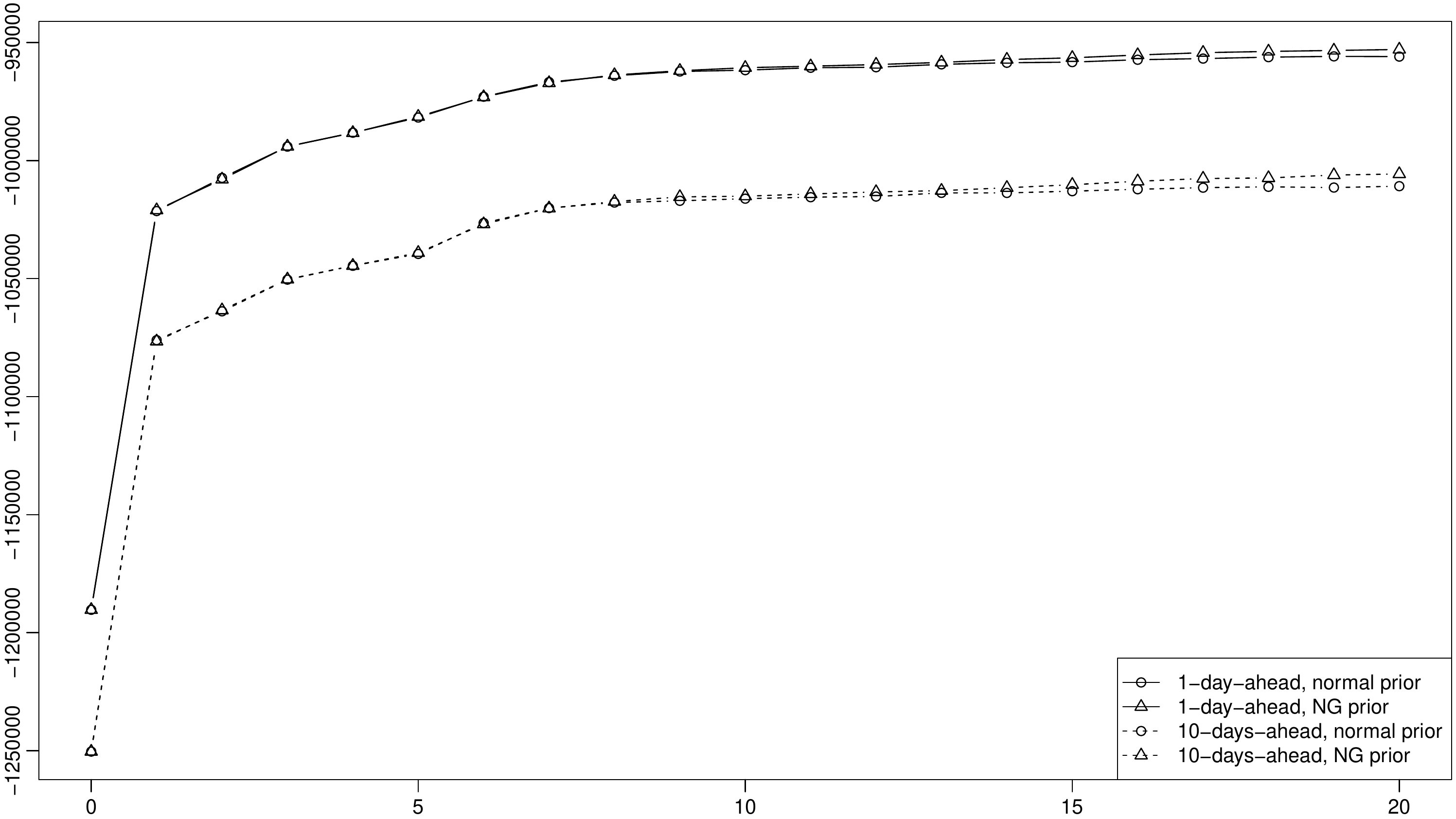}
\caption{Accumulated 1-day- and 10-days-ahead log predictive likelihoods for models with $0,1,\dots,20$ factors. Data until $t=3000$ is treated as training data.}
\label{pred_final}
\end{figure}

Accumulated log predictive likelihoods for the entire period are displayed in Figure~\ref{pred_final}. Gains in predictive power are substantial up to around 8 factors with little difference for the two priors. After this point, the benefit of adding even more factors turns out to be less pronounced. On the contrary, the effect of the priors becomes more pronounced. Again, while differences in scores tend to be muted for models with fewer factors, the benefit of shrinkage grows when $r$ gets larger.

While joint models with $r > 0$ outperform the marginal (no-factor) model for all points in time, days of particular turbulence particularly stand out. To illustrate this, we display average and top three log predictive gains over the no-factor model in Table~\ref{LPS1}. The biggest gains can be seen on ``Black Monday 2011'' (August 8), when US stock markets tumbled after a credit rating downgrade of US sovereign debt by Standard and Poor's. The trading day before this, August 4, displays the third highest gain. The 27th of February in 2007 also proves to be an interesting date to consider. This day corresponds to the burst of the Chinese stock bubble that led to a major crash in Chinese stock markets, causing a severe decline in equity markets worldwide.
It appears that joint modeling of stock prices is particularly important on days of extreme events when conditional correlations are often higher.

\include*{tables/LPS1ahead}

\subsection{Using Observable Instead of Latent Factors}
\label{sec:ff}

An alternative to estimating latent factors from data is to use observed factors instead \citep[cf.][]{wan-etal:dyn}. To explore this route, we investigate an alternative model with four observed factors, the three Fama-French plus the Momentum factor.\footnote{The Fama-French+Momentum factors are available at a daily frequency at Kenneth French's web page at \url{http://mba.tuck.dartmouth.edu/pages/faculty/ken.french/data_library.html}. The data was downloaded on February 21, 2017; missing values were replaced with zeroes and the data was standardized to have unconditional mean zero and variance one.} The Fama-French factors consist of the excess return on the market, a size factor SMB, and a book-to-market factor HML \citep{fam-fre:com}; the momentum factor MOM \citep{car:on} captures the empirically observed tendency for falling asset prices to fall further, and rising prices to keep rising. For estimation of this model we proceed exactly as before, except that we omit the last step of our posterior sampler and keep $\bm{f}$ fixed at the observed values.

Without presenting qualitative results in detail due to space constraints, we note that both the loadings on and the volatilities of the market excess returns show a remarkably close resemblance to those corresponding to the first latent factor displayed in Figures~\ref{sp500facload} and \ref{faclogvars}. To a certain extent (although much less pronounced), this is also true for SMB and the second latent factor as well as HML and the fourth latent factor. However, most of the loadings on MOM are shrunk towards zero and there is no recognizable similarity to the remaining latent factor from the original model. Log predictive scores for this model are very close to those for the SV model with two latent factors. In what follows, we term this approach FF+MOM.

\subsection{Comparison to Other Models}
\label{comparison}

We now turn to investigating the statistical performance of the factor SV model via out-of-sample predictive measures as well as its suitability for optimal asset allocation. The competitors are: Moving averages (MAs) of sample covariance matrices over a window of 500 trading days, exponentially weighted moving averages (EWMAs) of sample covariances defined by
$
\bm{\Sigma}_{t+1} = (1-\alpha)\bm{y}_t' \bm{y}_t + \alpha \bm{\Sigma}_{t},
$
the Ledoit-Wolf shrinkage estimator \citep{led-wol:hon} and FF+MOM described above. While this choice is certainly not exhaustive, it includes many of the approaches most widely used in practice. 

For comparison, we use two benchmarking methods.
First, we consider the minimum variance portfolio implied by $\bm{\hat\Sigma}_{t+1}$ (the point estimate or posterior mean estimate, respectively) which uniquely defines the optimal portfolio weights $$ \bm{\omega}_{t+1} = \frac{\bm{\hat\Sigma}_{t+1}^{-1}\bm{\iota}}{\bm{\iota}'\bm{\hat\Sigma}_{t+1}^{-1}\bm{\iota}}, $$ where $\bm{\iota}$ denotes an $m$-variate vector of ones. Using these weights, we compute the corresponding realized portfolio returns $r_{t+1}$ for $t = 3000,3001,\dots,3999$, effectively covering an evaluation period from 5/3/2006 to 3/1/2010. In the first three columns of Table \ref{GMVP}, we report annualized empirical standard deviations, annualized average excess returns over those obtained from the equal weight portfolio, and the quotient of these two measures, the Sharpe ratio \citep{sha:mut}.

\include*{tables/GMVP}

Considering the portfolio standard deviation presented in the first column of Table~\ref{GMVP}, it turns out that the Ledoit-Wolf shrinkage estimator implies an annualized standard deviation of about $12.5$ which is only matched by factor SV models with many factors. Lower-dimensional factor SV models, including FF+MOM, as well as simple MAs and highly persistent EWMAs do not perform quite as well but are typically well below 20. Less persistent EWMAs, the no-factor SV model and the na\"{i}ve equal weight portfolio exhibit standard deviations higher than 20. Considering average returns, FF+MOM and factor SV models with around 10 to 20 factors tend to do well for the given time span. In the third column, we list Sharpe ratios, where factor SV models with 10 to 20 factors show superior performance, in particular when the Normal-Gamma prior is employed. Note however that column two and three have to be interpreted with some care, as average asset and portfolio returns generally have a high standard error.

Second, we use what we coin \emph{pseudo log predictive scores} (PLPSs), i.e.\ Gaussian approximations to the actual log predictive scores. This simplification is necessary because most of the above-mentioned methods only deliver point estimates of the forecast covariance matrix and it is not clear how to properly account for estimation uncertainty. Moreover, the PLPS is simpler to evaluate as there is no need to numerically solve a high-dimensional integral. Consequently, it is frequently used instead of the actual LPS in high dimensions while still allowing for evaluation of the covariance accuracy \citep{ado-etal:for,car-etal:com,hub:den}. More specifically, we use data up to time $t$ to determine a point estimate $\bm{\hat\Sigma}_{t+1}$ for $\bm{\Sigma}_{t+1}$ and compute the logarithm of the multivariate Gaussian density $\Normult{m}{\bfz, \bm{\hat\Sigma}_{t+1}}$ evaluated at the actually observed value $\bm{y}^o_{t+1}$ to obtain the one-day-ahead PLPS for time $t + 1$.

In terms of average PLPSs (the last column in Table~\ref{GMVP}), factor SV models clearly outperform all other models under consideration. In particular, even if $r$ is chosen as small as $r=4$ to match the number of factors, the model with latent factors outperforms FF+MOM. Using latent factors is generally preferable; note however that the 4-factor FF+MOM does better than the single- and no-factor SV models. Generally speaking, many factors appear to be needed for accurately representing the underlying data structure, irrespectively of the prior choice. Considering the computational simplicity of the Ledoit-Wolf estimator, its prediction accuracy is quite remarkable. It clearly outperforms the no-factor SV model which, in turn, beats simple MAs and EWMAs.

\section{Conclusion and Outlook}
\label{sec:concl}

The aim of this paper was to present an efficient and parsimonious method of estimating high-dimensional time-varying covariance matrices through factor stochastic volatility models. We did so by proposing an efficient Bayesian MCMC algorithm that incorporates parsimony by modeling the covariance structure through common latent factors which themselves follow univariate SV processes. Moreover, we added additional sparsity by utilizing a hierarchical shrinkage prior, the Normal-Gamma prior, on the factor loadings. We showed the effectiveness of our approach through simulation studies and illustrated the effect of different shrinkage specifications. We applied the algorithm to a high-dimensional data set consisting of stock returns of $300$ S\&P 500 members and conducted an out-of-sample predictive study to compare different prior settings and investigate the choice of the number of factors. Moreover, we discussed the out-of-sample performance of a minimum variance portfolio constructed from the model-implied weights and related it to a number of competitors often used in practice.

Because the algorithm scales linearly in both the series length $T$ as well as the number of component series $m$, applying it to even higher dimensions is straightforward. We have experimented with simulated data in thousands of dimensions for thousands of points in time and successfully recaptured the time-varying covariance matrix.

Further research could be directed towards incorporating prior knowledge into building the hierarchical structure of the Normal-Gamma prior, e.g.~by choosing the global shrinkage parameters according to industry sectors. Alternatively, \cite{vil-etal:reg} propose a mixture of experts model to cater for smoothly changing regression densities. It might be fruitful to adopt this idea in the context of covariance matrix estimation by including either observed (Fama-French) or latent factors as predictors and allowing for other mixture types than the ones discussed there.
While not being the focus of this work, it is easy to extend the proposed method by exploiting the modular nature of Markov chain Monte Carlo methods. In particular, it is straightforward to combine it with mean models such as (sparse) vector autoregressions \citep[e.g.,][]{ban-etal:lar, kas-hub:spa}, dynamic regressions \citep[e.g.,][]{kor:hie}, or time-varying parameter models \citep[e.g.,][]{koo-kor:lar, hub-etal:new}.


\section*{Acknowledgments}
Variants of this paper were presented at the 6th European Seminar on Bayesian Econometrics (ESOBE 2015), the 2015 NBER-NSF Time Series Conference, the 2nd Vienna Workshop on High-Dimensional Time Series in Macroeconomics and Finance 2015, the 2015 International Work-Conference on Time Series Analysis, the 2016 ISBA World Meeting, the 10th International Conference on Computational and Financial Econometrics 2016, the CORE Econometrics \& Finance Seminar 2017 and the 61st World Statistics Congress 2017. The author thanks all participants, in particular Luc Bauwens, Manfred Deistler, John Geweke, Florian Huber, Sylvia Kaufmann, Hedibert Lopes, Ruey Tsay, Stefan Voigt, Mike West, as well as the handling editor Herman van Dijk and two anonymous referees for crucially valuable comments and suggestions. Special thanks go to Mark Jensen who discussed this paper at the ESOBE 2015 and Sylvia Fr\"uhwirth-Schnatter who continuously supported the author throughout the development of this work.

\singlespacing
\small
\printbibliography[heading = bibintoc]

\end{document}

%% file: macros.tex
\newcommand{\pkg}[1]{{\normalfont\fontseries{b}\selectfont #1}}

\let\proglang=\textsf

\newcommand{\allpara}{\bm{\theta}}
\newcommand{\unobs}{\bm{\kappa}}

\newcommand{\yb}{\bm{y}}

\newcommand{\m}{k}
\newcommand{\M}{K}

\newcommand{\Gammad}[1]{ \mathcal{G}\!\left(#1\right)}
\newcommand{\Normal}[1]{\mathcal{N}\!\left(#1\right)}

\newcommand{\Normult}[2]{\mathcal{N}_{#1}\!\left(#2\right)}

\newcommand{\Betad}[1]{\mathcal{B}\!\left(#1\right)}

\newcommand{\ym}{{\bm{ y}}}             

\newcommand{\fm}{{\bm{ f}}}             
\newcommand{\Vary}{{\bm{\vary}}}         
\newcommand{\vary}{\Sigma}         
\newcommand{\dimy}{m}                     
\newcommand{\load}{\Lambda}                 
 
\newcommand{\facload}{\bm{\load}} 

\newcommand{\fac}{f}                      
\newcommand{\facm}{{\bm{ \fac}}}        

\newcommand{\Vare}{{\bm{\vare}}}      
\newcommand{\vare}{{{U}}}      
\newcommand{\hsv}{h}             
\newcommand{\hsvm}{{\bm{\hsv}}}             

\newcommand{\Vm}{{\bm{ V}}}             

\newcommand{\mupar}{\mu}
\newcommand{\phipar}{\phi}
\newcommand{\sigmapar}{\sigma}

\newcommand{\Diag}[1]{\mbox{\text diag}\!\left(#1\right)} 
\newcommand{\bfz}{{\bm{{0}}}}         
\newcommand{\trans}[1]{#1 ^{'}}         

\newcommand{\nfactrue}{r}                     

\theoremstyle{plain}

\theoremstyle{definition}

\newcommand{\e}{e}

\newcommand{\dif}{\mathrm{d}}

\newcommand{\PL}{P\!L}
\newcommand{\BF}{B\!F}

%% file: tables/predictive_sim_10/tables.tex
\begin{table}[t]
\centering
\begin{tabular}{lrrrrrr}
  \hline
 & no fac & 1 fac & 2 fac & 3 fac & 4 fac & 5 fac \\ 
  \hline
Standard Gaussian & $-$845.80 & $-$316.41 &  & $-$2.81 & $-$5.83 & $-$8.38 \\ 
  Row-wise NG ($c_i=d_i=1$) & $-$845.80 & $-$317.75 & 0.27 & $-$0.44 & $-$1.74 & $-$2.34 \\ 
  Col-wise NG ($c_j=d_j=1$) & $-$845.80 & $-$317.45 & 0.38 & $-$0.77 & $-$1.61 & $-$2.06 \\ 
  Row-wise NG ($c_i=d_i=0.01$) & $-$845.80 & $-$317.71 & 0.40 & $-$0.29 & $-$1.36 & $-$1.97 \\ 
  Col-wise NG ($c_j=d_j=0.01$) & $-$845.80 & $-$317.52 & 1.04 & $-$0.12 & $-$0.88 & $-$2.07 \\ 
   \hline
Standard Gaussian & $-$787.41 & $-$255.92 &  & $-$4.98 & $-$7.75 & $-$12.67 \\ 
  Row-wise NG ($c_i=d_i=1$) & $-$787.41 & $-$256.85 & 2.90 & 1.73 & 0.87 & $-$0.05 \\ 
  Col-wise NG ($c_j=d_j=1$) & $-$787.41 & $-$256.44 & 3.14 & 1.17 & 0.90 & 0.09 \\ 
  Row-wise NG ($c_i=d_i=0.01$) & $-$787.41 & $-$256.81 & 3.18 & 1.98 & 1.28 & 0.53 \\ 
  Col-wise NG ($c_j=d_j=0.01$) & $-$787.41 & $-$255.90 & 2.85 & 1.61 & 0.84 & 0.16 \\ 
   \hline
\end{tabular}
\caption{Estimated log Bayes factors at $t=1500$ against the 2-factor model using a standard Gaussian prior, where
	     data up to $t=1000$ is treated as the training sample. Lines 1 to 5 correspond to cumulative 1-day-ahead Bayes factors,
	     lines 6 to 10 correspond to 10-days-ahead predictive Bayes factors.} 
\label{pred10}
\end{table}

%% file: tables/10/rmses_per_series_rel_small.tex
\begin{table}[t]
\centering
\begin{tabular}{rrrrrrrrrrr}
  & 1 & 2 & 3 & 4 & 5 & 6 & 7 & 8 & 9 & 10 \\ 
  \hline
Average 1 & 1.05 & 1.03 & 1.05 & 1.07 & 1.04 & 1.05 & 1.08 & 1.03 & 1.32 & 1.05 \\ 
   \hline
1 &  & 1.00 & 1.00 & 1.08 & 1.00 & 1.00 & 1.10 & 1.00 & 1.29 & 1.00 \\ 
  2 & 1.00 &  & 1.00 & 1.00 & 1.00 & 1.00 & 1.00 & 1.00 & 1.26 & 1.00 \\ 
  3 & 1.00 & 1.01 &  & 1.07 & 1.00 & 1.00 & 1.09 & 1.00 & 1.31 & 0.99 \\ 
  4 & 3.04 & 1.00 & 2.85 &  & 1.06 & 1.07 & 1.00 & 1.00 & 1.30 & 1.10 \\ 
  5 & 1.00 & 1.01 & 1.00 & 3.01 &  & 1.00 & 1.06 & 1.00 & 1.31 & 1.00 \\ 
  6 & 1.00 & 1.01 & 1.00 & 3.00 & 1.00 &  & 1.07 & 1.00 & 1.32 & 0.99 \\ 
  7 & 3.02 & 1.00 & 2.81 & 1.00 & 2.87 & 2.77 &  & 1.00 & 1.28 & 1.09 \\ 
  8 & 1.00 & 1.00 & 1.00 & 1.01 & 1.01 & 1.01 & 1.01 &  & 1.27 & 1.00 \\ 
  9 & 2.85 & 2.61 & 2.70 & 2.69 & 2.73 & 2.71 & 2.77 & 2.54 &  & 1.31 \\ 
  10 & 1.00 & 1.01 & 1.00 & 2.84 & 1.00 & 1.00 & 2.89 & 1.01 & 2.72 &  \\ 
   \hline
Average 2 & 1.45 & 1.12 & 1.40 & 2.01 & 1.40 & 1.41 & 2.04 & 1.12 & 2.79 & 1.40 \\ 
   \hline
\end{tabular}
\caption{Relative RMSEs of pairwise correlations. Above the diagonal: Row-wise Lasso ($a_i=1$) vs.\ benchmark standard Gaussian prior with geometric means (first row). Below the diagonal: Row-wise Normal-Gamma ($a_i=0.1$) vs.\ row-wise Lasso prior ($a_i=1$) with geometric means (last row). Numbers greater than one mean that the former prior performs better than the latter. Hyperhyperparameters are set to $c_i=d_i=0.001$. All values reported are medians of 100 repetitions.} 
\label{rmsessmall}
\end{table}

%% file: tables/10/rmses_per_series_rel_large.tex
\begin{table}[t]
\centering
\begin{tabular}{rrrrrrrrrrr}
  & 1 & 2 & 3 & 4 & 5 & 6 & 7 & 8 & 9 & 10 \\ 
  \hline
Average 1 & 1.00 & 1.00 & 1.01 & 1.01 & 1.01 & 1.01 & 1.01 & 1.00 & 1.02 & 1.00 \\ 
   \hline
1 &  & 1.00 & 1.00 & 1.01 & 1.00 & 1.00 & 1.01 & 1.00 & 1.01 & 1.00 \\ 
  2 & 1.00 &  & 1.00 & 1.00 & 1.00 & 1.00 & 1.00 & 1.00 & 1.01 & 1.00 \\ 
  3 & 1.00 & 1.01 &  & 1.01 & 1.00 & 1.00 & 1.02 & 1.00 & 1.03 & 1.00 \\ 
  4 & 3.06 & 1.00 & 2.99 &  & 1.02 & 1.02 & 1.00 & 1.00 & 1.01 & 1.02 \\ 
  5 & 1.00 & 1.01 & 1.00 & 3.02 &  & 1.00 & 1.02 & 1.00 & 1.03 & 1.00 \\ 
  6 & 1.00 & 1.01 & 1.00 & 2.93 & 1.00 &  & 1.02 & 1.00 & 1.04 & 1.00 \\ 
  7 & 3.25 & 1.00 & 2.98 & 1.00 & 2.90 & 2.72 &  & 1.00 & 1.02 & 1.02 \\ 
  8 & 1.00 & 1.00 & 1.00 & 1.01 & 1.01 & 1.00 & 1.01 &  & 1.02 & 1.00 \\ 
  9 & 3.51 & 3.07 & 3.28 & 3.19 & 3.40 & 3.37 & 3.28 & 3.03 &  & 1.02 \\ 
  10 & 1.00 & 1.01 & 1.00 & 2.92 & 1.00 & 1.00 & 2.98 & 1.01 & 3.32 &  \\ 
   \hline
Average 2 & 1.49 & 1.14 & 1.43 & 2.11 & 1.43 & 1.42 & 2.07 & 1.14 & 3.37 & 1.44 \\ 
   \hline
\end{tabular}
\caption{Relative RMSEs of pairwise correlations. Above the diagonal: Row-wise Lasso ($a_i=1$) vs.\ benchmark standard Gaussian prior with geometric means (first row). Below the diagonal: Row-wise Normal-Gamma ($a_i=0.1$) vs.\ row-wise Lasso prior ($a_i=1$) with geometric means (last row). Numbers greater than one mean that the former prior performs better than the latter. Hyperhyperparameters are set to $c_i=d_i=1$. All values reported are medians of 100 repetitions.} 
\label{rmseslarge}
\end{table}

%% file: tables/10/rmses_per_series_overall.tex
\begin{table}[t]
\centering
\begin{tabular}{rrrrrr}
  \hline
 & GSH & Abs. RMSE & Rel. RMSE & Abs. MAE & Rel. MAE \\ 
  \hline
Standard Gaussian &  & 7.581 &  & 5.397 &  \\ 
   \hline
Row-wise Lasso & 0.001 & 7.587 & 100.449 & 5.324 & 101.850 \\ 
  Col-wise Lasso & 0.001 & 7.593 & 100.176 & 5.372 & 100.622 \\ 
  Row-wise Lasso  & 1.000 & 7.602 & 99.995 & 5.416 & 100.146 \\ 
  Col-wise Lasso  & 1.000 & 7.584 & 99.987 & 5.415 & 100.119 \\ 
   \hline
Row-wise NG & 0.001 & 7.351 & 102.793 & 4.891 & 110.252 \\ 
  Col-wise NG & 0.001 & 7.395 & 102.525 & 4.957 & 109.912 \\ 
  Row-wise NG  & 1.000 & 7.392 & 102.641 & 4.941 & 109.735 \\ 
  Col-wise NG  & 1.000 & 7.384 & 102.622 & 4.925 & 109.722 \\ 
   \hline
\end{tabular}
\caption{Different error measures ($\times 10^{-2}$) of posterior mean correlation estimates under various priors. For all scenarios, $3$-factor SV models are fit to $10$-dimensional data of length $1000$ simulated from a $2$-factor SV model. The column titled ``GSH'' contains the values of the global shrinkage hyperhyperparameters $c_i=c_j=d_i=d_j$. All values reported are medians of 100 repetitions.} 
\label{rmsesoverall}
\end{table}

%% file: tables/100/rmses_per_series_overall_cor.tex
\begin{table}[t]
\centering
\begin{tabular}{rrrrrr}
  \hline
 & GSH & Abs. RMSE & Rel. RMSE & Abs. MAE & Rel. MAE \\ 
  \hline
Standard Gaussian &  & 8.906 &  & 7.074 &  \\ 
   \hline
Row-wise Lasso & 0.001 & 8.293 & 107.047 & 6.553 & 107.844 \\ 
  Col-wise Lasso & 0.001 & 8.425 & 105.642 & 6.659 & 106.125 \\ 
  Row-wise Lasso  & 1.000 & 8.215 & 107.811 & 6.496 & 108.392 \\ 
  Col-wise Lasso  & 1.000 & 8.345 & 106.508 & 6.609 & 107.058 \\ 
   \hline
Row-wise NG & 0.001 & 7.773 & 114.466 & 6.074 & 116.411 \\ 
  Col-wise NG & 0.001 & 7.802 & 114.525 & 6.085 & 116.768 \\ 
  Row-wise NG  & 1.000 & 7.799 & 114.386 & 6.099 & 116.391 \\ 
  Col-wise NG  & 1.000 & 7.778 & 114.121 & 6.070 & 116.410 \\ 
   \hline
\end{tabular}
\caption{Different error measures ($\times 10^{-2}$) of posterior mean correlation estimates under various priors. For all scenarios, unrestricted $11$-factor SV models are fit to $100$-variate data of length $1000$ simulated from $10$-factor SV models. The column titled ``GSH'' contains the values of the global shrinkage hyperhyperparameters $c_i=c_j=d_i=d_j$. All values reported are medians of 100 repetitions.} 
\label{rmsesoverall100}
\end{table}

%% file: tables/LPS1ahead.tex
\begin{table}[t]
\centering
\begin{tabular}{rrrrr}
  \hline
 & Mean & 2/27/2007 & 8/4/2011 & 8/8/2011 \\ 
  \hline
1 factors, Gaussian & 84.83 & 1415.13 & 1189.78 & 1446.03 \\ 
  2 factors, Gaussian & 91.88 & 1455.39 & 1261.38 & 1462.23 \\ 
  3 factors, Gaussian & 98.55 & 1449.31 & 1237.97 & 1479.84 \\ 
  4 factors, Gaussian & 101.50 & 1449.21 & 1254.36 & 1502.17 \\ 
  10 factors, Gaussian & 114.81 & 1482.36 & 1271.59 & 1557.63 \\ 
  20 factors, Gaussian & 117.69 & 1481.15 & 1293.98 & 1568.60 \\ 
   \hline
1 factors, NG prior & 85.03 & 1421.45 & 1194.08 & 1445.51 \\ 
  2 factors, NG prior & 91.56 & 1450.88 & 1239.42 & 1461.73 \\ 
  3 factors, NG prior & 98.57 & 1448.90 & 1231.94 & 1471.58 \\ 
  4 factors, NG prior & 101.47 & 1443.86 & 1240.68 & 1536.77 \\ 
  10 factors, NG prior & 115.32 & 1482.32 & 1283.90 & 1546.43 \\ 
  20 factors, NG prior & 119.21 & 1499.08 & 1293.74 & 1578.54 \\ 
   \hline
\end{tabular}
\caption{Average and top 3 daily 1-day-ahead log predictive gains over the no-factor SV model.} 
\label{LPS1}
\end{table}

%% file: tables/GMVP.tex
\begin{table}[t]
\centering
\begin{tabular}{rrrrr}
  \hline
 & SD & Avg & Sharpe & PLPS \\ 
  \hline
Equal weight portfolio & 29.02 & 0.00 & 0.00 &  \\ 
  MA (500 days) & 16.56 & $-$2.39 & $-$0.14 & $-$320.17 \\ 
  EWMA ($\alpha = 0.94$) & 35.90 & $-$4.43 & $-$0.12 & $-3\times10^{9}$ \\ 
  EWMA ($\alpha = 0.99$) & 18.42 & $-$4.18 & $-$0.23 & $-$1409.97 \\ 
  Ledoit-Wolf & 12.53 & 6.14 & 0.49 & 25.68 \\ 
  FF+MOM & 16.82 & 12.68 & 0.75 & 74.56 \\ 
  no-factor SV & 22.53 & 1.02 & 0.05 & 0.00 \\ 
  N-FSV 1 & 19.03 & 9.45 & 0.50 & 66.29 \\ 
  N-FSV 2 & 18.92 & 9.32 & 0.49 & 75.25 \\ 
  N-FSV 3 & 15.50 & 7.36 & 0.48 & 81.47 \\ 
  N-FSV 4 & 14.58 & 8.26 & 0.57 & 87.07 \\ 
  N-FSV 10 & 12.94 & 12.45 & 0.96 & 97.74 \\ 
  N-FSV 20 & 12.57 & 7.81 & 0.62 & 101.82 \\ 
  N-FSV 50 & 12.41 & 5.03 & 0.40 & 106.57 \\ 
  NG-FSV 1 & 18.98 & 9.44 & 0.50 & 66.29 \\ 
  NG-FSV 2 & 18.84 & 9.37 & 0.50 & 75.21 \\ 
  NG-FSV 3 & 15.52 & 7.20 & 0.46 & 81.39 \\ 
  NG-FSV 4 & 14.85 & 8.36 & 0.56 & 86.80 \\ 
  NG-FSV 10 & 12.94 & 12.00 & 0.93 & 97.74 \\ 
  NG-FSV 20 & 12.02 & 13.33 & 1.11 & 102.30 \\ 
  NG-FSV 50 & 12.42 & 10.78 & 0.87 & 107.23 \\ 
   \hline
\end{tabular}
\caption{Predictive performance measures, averaged over 1000 trading days after 5/3/2006. SD: Annualized empirical standard deviations of portfolio returns. Avg: Annualized average excess returns over the equal weight portolio. Sharpe: Quotients of Avg and SD. PLPS: Average one-day-ahead pseudo log predictive scores over the no-factor SV model. N-FSV stands for the factor SV model with the standard normal prior. NG-FSV stands for the factor SV model with row-wise Normal-Gamma prior ($a_i \equiv 0.1$).}
\label{GMVP}
\end{table}